\documentclass[useAMS,usenatbib,usegraphicx]{mn2e}

\usepackage{epsf,amsmath,rotating}
\def\apj{ApJ}
\def\apjl{ApJL}

\def\aj{AJ}
\def\mnras{MNRAS}
\def\aa{A\&A}
\def\na{NewA}
\def\phr{Phys Rep.}
\def\pasj{PASJ}
\def\pasp{PASP}

\newcommand{\bq}{\begin{equation}}
\newcommand{\eq}{\end{equation}}
\newcommand{\bqn}{\begin{eqnarray}}
\newcommand{\eqn}{\end{eqnarray}}

\newcommand{\dd}{\mbox{\rm d}}

\newcommand{\msun}{\rm{M}_\mathrm{\rm \sun}}

\def\etal{{\it et al.}}
\title[Clumps in the tidal tails of star clusters]{\sc Quantitative analysis 
of clumps in the tidal tails of star clusters}

\author[A. Just et al.]{A. Just$^1$, P. Berczik$^{1,2}$, M.I. Petrov$^{3,2}$, 
A. Ernst$^{1,4}$.\\
  $^{1}$ Astronomisches Rechen-Institut, Zentrum f\"{u}r Astronomie der 
  Universit\"{a}t Heidelberg (ZAH), M\"{o}nchhofstra\ss{}e 12-14, 69120 
  Heidelberg, Germany\\
  $^{2}$ Main Astronomical Observatory, National Academy of Sciences of Ukraine, 
  27 Akademika Zabolotnoho St., 03680 Kyiv, Ukraine\\
  $^{3}$ Institut f\"{u}r Astronomie der Universit\"{a}t Wien, 
  T\"{u}rkenschanzstra\ss{}e 17, A-1180 Wien, Austria\\
  $^{4}$ Max-Planck-Institut f\"ur Astronomie, K\"onigstuhl 17, 
  69117 Heidelberg, Germany}

\begin{document} 

\date{Accepted 2008 August xx. Received 2008 August yy; in original form 2008 August zz}

\pagerange{\pageref{firstpage}--\pageref{lastpage}} \pubyear{2008} 

%%%%%%%%%%%%%%%%%%%%%%%%%%%%%%%%%%%%%%%%%%%%%%%%%%%%%%%%%%%%%%%%%%%%%%%%%%%%%

\maketitle

\label{firstpage} 

%%%%%%%%%%%%%%%%%%%%%%%%%%%%%%%%%%%%%%%%%%%%%%%%%%%%%%%%%%%%%%%%%%%%%%%%%%%%%

\begin{abstract}
Tidal tails of star clusters are not homogeneous but show well 
defined clumps in observations as well as in numerical 
simulations. Recently an epicyclic theory for the formation of 
these clumps was presented. A quantitative analysis was still 
missing. We present a quantitative derivation of the angular 
momentum and energy distribution of escaping stars from a star 
cluster in the tidal field of the Milky Way and derive the 
connection to the position and width of the clumps. For the 
numerical realization we use star-by-star $N$-body simulations. 
We find a very good agreement of theory and models. 
We show that the radial offset of the tidal arms scales with the 
tidal radius, which is a function of cluster mass and the rotation 
curve at the cluster orbit. The mean radial
offset is 2.77 times the tidal radius in the outer disc.
Near the Galactic centre the circumstances are more complicated, but to lowest
order the theory still applies. We have also measured the Jacobi 
energy distribution of bound stars and showed that there is a 
large fraction of stars (about 35\%) above the critical Jacobi 
energy at all times, which can potentially leave the cluster. 
This is a hint that the mass loss is dominated by a self-regulating 
process of increasing Jacobi energy due to the weakening of 
the potential well of the star cluster, which is induced by the 
mass loss itself. 
\end{abstract}

%%%%%%%%%%%%%%%%%%%%%%%%%%%%%%%%%%%%%%%%%%%%%%%%%%%%%%%%%%%%%%%%%%%%%%%%%%%%%


%%%%%%%%%%%%%%%%%%%%%%%%%%%%%%%%%%%%%%%%%%%%%%%%%%%%%%%%%%%%%%%%%%%%

%Version: tidal-clumps7; \today

%%%%%%%%%%%%%%%%%%%%%%%%%%%%%%%%%%%%%%%%%%%%%%%%%%%%%%%%%%%%%%%%%%%%%%%%%%%%%

\section{Introduction}
\label{sec-intro}

Recently well-defined clumps were observed in the tidal tails of globular
clusters by \citet{Le00} for NGC 6254 and Pal 12 and by
\citet{Od01,Od03} for Pal 5. 
External perturbations like crossing of the galactic disc, peri-centre passage
or near-encounters with other globular clusters were discussed as the source of
the clumps \citep{Ca05}. In \citet{Ca05} the formation of these clumps in tidal
tails of star clusters on eccentric orbits were 
confirmed by numerical studies. But the clumps 
occur also in the tidal tails of star clusters on 
circular orbits with no external push.
Recently, \citet{Ku08} presented a theoretical explanation for
the clump formation in a constant tidal field. 
It is essentially due to the epicyclic motion of the stars
lost by the star cluster. 

We present a quantitative analysis of the tidal tail structure for star
clusters moving on a circular orbit in the galactic disc. The analysis is based
on numerical simulations with realistic particle numbers including an initial
mass function (IMF) and stellar evolution. We compare the results for star
clusters at the solar circle and near the Galactic centre.
Since the epicycle theory is a perturbation theory with respect to a circular
orbit with constant tidal field, we cannot apply it for predictions of clump
distances to the eccentric orbits of
the observed globular clusters.
On the other hand the observation of tidal tail clumps of open clusters in the
galactic disc are hampered by the overwhelming number of field stars with
similar properties. For an identification the contrast in density and
velocity with respect to the field stars may
be too small. Additionally the tidal tails may be destroyed quickly by the same
gravitational scattering process, which is also responsible for the
dynamical heating of the stellar disc. We discuss the observability further in
Sect.~\ref{sec-sum}.

In Section \ref{sec-dyn} we present the epicyclic theory for the stars
in the tidal tails and the connection to the mass loss and the orbit of the 
star cluster. In Section \ref{sec-num} we present the numerical codes 
used and the properties of the star clusters. Section \ref{sec-res} 
contains the quantitative comparison of the numerical results with the 
theoretical predictions. In Section \ref{sec-sum} we summarize our results.

%%%%%%%%%%%%%%%%%%%%%%%%%%%%%%%%%%%%%%%%%%%%%%%%%%%%%%%%%%%%%%%%%%%%%%%%%%%%%

\section{Dynamics of escaping stars}
\label{sec-dyn}

In the first part of this section we derive the orbital properties of the stars
in the tidal tails in terms of angular momentum and energy. Then we discuss 
the connection of mass loss with the Jacobi energy distribution of the stars.
Finally the tidal tail properties are determined with respect to the star
cluster orbit.

Classically the gravitational potential $\Phi_\mathrm{g}(R)$ and 
kinetic energy $\Omega^2 R^2/2$ are approximated to second order in
the variable $r=R-R_0$, which we will also use. 
For the determination
of the tidal tail properties with respect to the star cluster orbit in 
Section~\ref{sec-tid} we switch to 
the Taylor expansion with respect to angular momentum 
(e.g. for $R_0(L)=\sqrt{L/\Omega(L)}$ equation~\ref{eq-RofL}), because $L$ is
a constant of motion for the tidal tail stars and easily measurable.

%%%%%%%%%%%%%%%%%%%%%%%%%%%%%%%%%%%%%%%%%%%%%%%%%%%%%%%%%%%%%%%%%%%%%%%%%%%%%

\subsection{Motion of tidal tail stars}
\label{sec-epi}

As soon as the gravitational potential of the star cluster is negligible, stars
in the tidal tails move in the axisymmetric potential of the Galaxy. 
The orbits can be calculated in the epicyclic approximation (see \citet{Ku08} for a
recent application to tidal tail structure). 
The radial offset of the epicyclic motion relative to the orbit of the star
cluster is first order in the angular momentum. The radial amplitude depends on
the energy excess, which is of second order.

%----------------------------------------------------------------------------%
\begin{figure}
  \begin{center}
% \vspace{-0.2\textwidth}
  \includegraphics[width=0.48\textwidth]{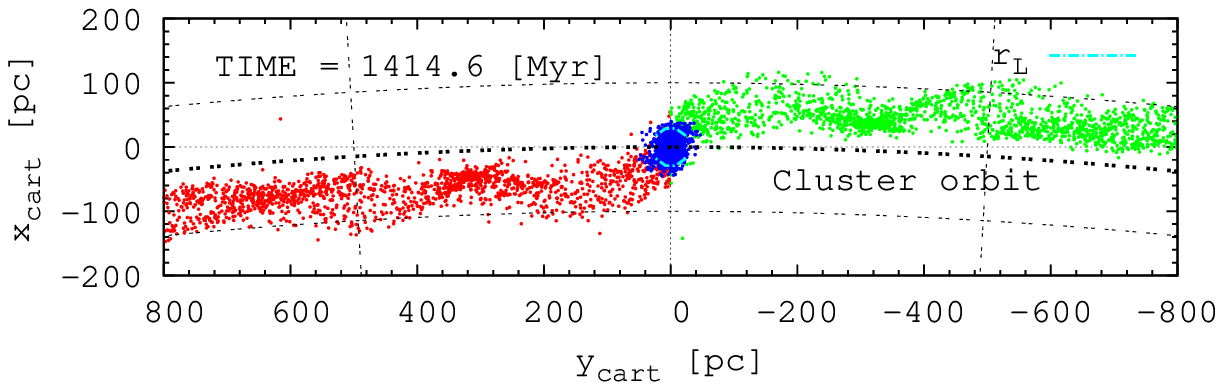}
  \includegraphics[width=0.48\textwidth]{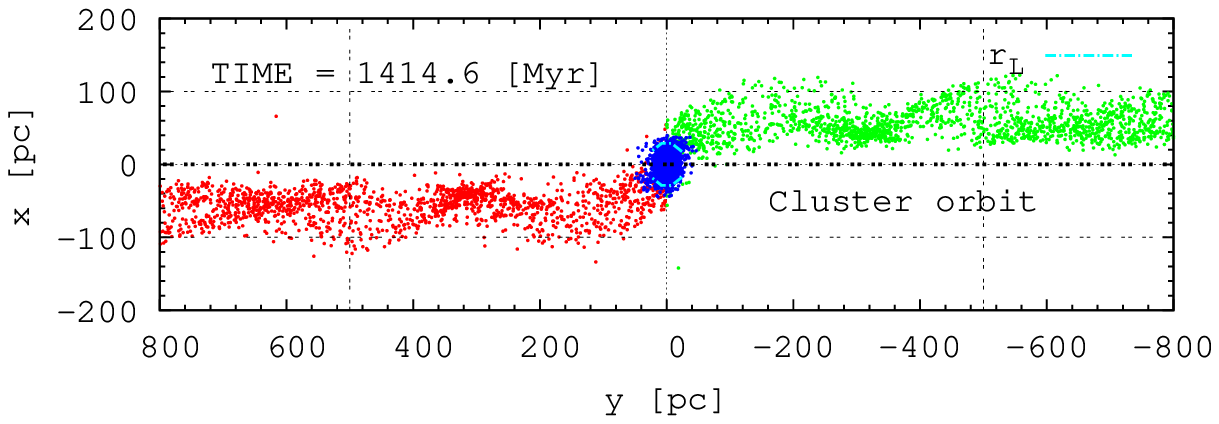}
  \end{center}
  \caption{Visualization of the local cartesian (top) and local cylindrical 
  coordinates (bottom) for the snapshot at t=1.414\,Gyr of model 10, where 
  the origin is
  moved from the galactic centre to the star cluster centre. 
  Black (blue) dots are the bound
  stars, dark grey dots to the left (red) are the stars of the leading arm and 
  light grey dots to the right (green) are the stars in the trailing
  arm. The light dashed (cyan) circle marks the tidal radius. }
  \label{pos-300}
\end{figure}
%----------------------------------------------------------------------------%

Since the tidal tails may extend
over a considerable range in azimuth, we use polar coordinates
$R,\varphi,z$ with origin at the galactic centre. 
Fig.~\ref{pos-300} shows the effect of switching from local cartesian
($x_\mathrm{cart},y_\mathrm{cart}$) to local polar ($x,y$) coordinates.
We restrict the investigation
to orbits in the galactic plane $z=0$, but a generalization is straightforward.
 Energy $E$ and the z-component of
angular momentum $L=L_{z}$ are isolating integrals of motion in the axisymmetric
potential of the Galaxy.
The motion is a 2-dimensional harmonic oscillation
with the epicyclic frequency $\kappa$ (see equation~\ref{eq-kappa}).
We use the normalized epicyclic frequency $\beta=\kappa/\Omega$.
 The 'epicentre'  (guiding centre) of the oscillation is described by
 \bq
(R,\varphi)_0=(R_0,\Omega_0 t)
\eq
where $\Omega(R)$ is the angular frequency of the galactic rotation and 
$\Omega_0=\Omega(R_0)$.
$R_0$ is determined by the angular momentum of the star via
\bq
L=L_0=\Omega_0 R_0^2
\eq
In one  epicyclic period $T=2\pi/\kappa_0$ the epicentre moves along 
the circle with radius $R_0$ by 
\bq
D_0(T)=\frac{2\pi}{\kappa_0}\Omega_0 R_0=\frac{2\pi}{\beta_0} R_0
\eq

The energy of the star determines the amplitude of the oscillation. At the apo-
and pericentre $R_\mathrm{m}$, where the radial motion vanishes, it can be written as
\bq
E=\Phi_\mathrm{g}(R_\mathrm{m})+\frac{L_0^2}{2 R_\mathrm{m}^2}
\eq
with the galactic potential $\Phi_\mathrm{g}(R)$.
Relative to the circular motion with
\bq
E_0=\Phi_\mathrm{g}(R_0)+\frac{L_0^2}{2 R_0^2}
\label{eq-e0}
\eq
the radial amplitude $r_\mathrm{m}=R_\mathrm{m}-R_0$ is determined by the energy excess
\bq
\Delta E=E-E_0=\Phi_\mathrm{g}(R_\mathrm{m})-\Phi_\mathrm{g}(R_0)+
	\frac{L_0^2}{2}\left(\frac{1}{R_\mathrm{m}^2}-\frac{1}{R_0^2}\right)
	\label{eq-deltae}
\eq
To second order in $r$ (see App.~\ref{app-epi}) we find for
the radial amplitude $r_\mathrm{m}$ and peri/apocentre position $R_\mathrm{m}$
\bqn
R_\mathrm{m}&=&R_0\pm r_\mathrm{m}
	=R_0\pm \frac{\sqrt{2 \Delta E}}{\beta_0 \Omega_0}
\eqn
Note that the epicentre is determined by the angular momentum $L$ of the star
and the radial amplitude by the energy excess $\Delta E$.

The amplitude in tangential direction is 
\bq
y_\mathrm{m}=R_0\Delta\varphi_\mathrm{m}=\frac{2}{\beta_0}r_\mathrm{m}
\eq
The epicyclic ellipse is elongated into the radial
direction, because $1\le \beta \le 2$ for reasonable rotation curves.

%%%%%%%%%%%%%%%%%%%%%%%%%%%%%%%%%%%%%%%%%%%%%%%%%%%%%%%%%%%%%%%%%%%%%%%%%%%%%

\subsection{Mass loss of the star cluster}
\label{sec-mass}

We investigate the mass loss of a star cluster on a circular orbit with
$R_\mathrm{C}, \Omega_\mathrm{C}$ in the tidal field of the Galaxy. The orbit of
the cluster is offset to the epicentres $R_0,\Omega_0$ of the stars in the 
tidal tails. The differential rotation results in the elongation of the tidal
tails.

%----------------------------------------------------------------------------%
\begin{figure}
%\begin{center}
\vspace{-0.9cm}
\hspace*{-0.8cm}\includegraphics[width=10cm, angle=0]{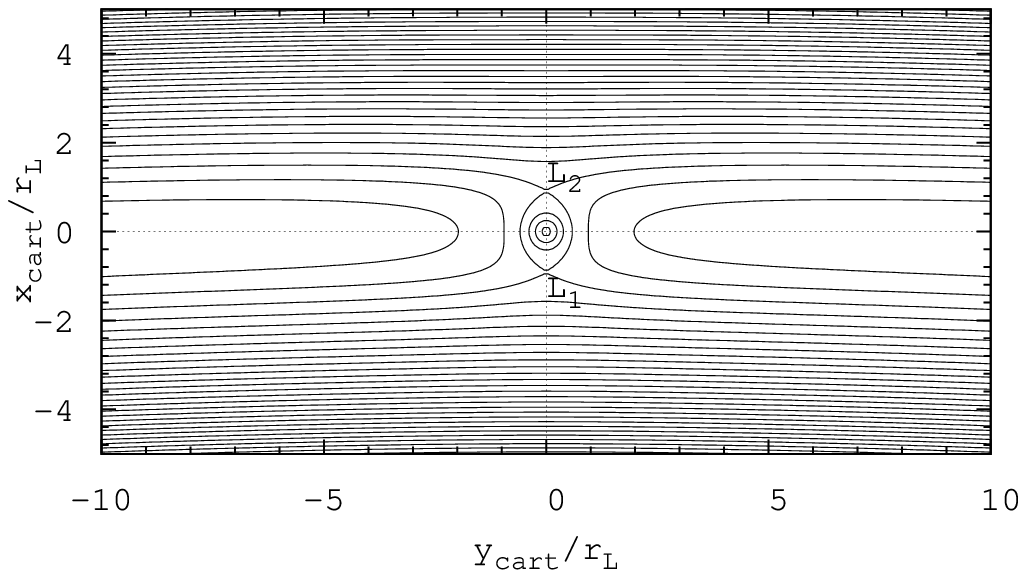}
\includegraphics[width=8cm, angle=0]{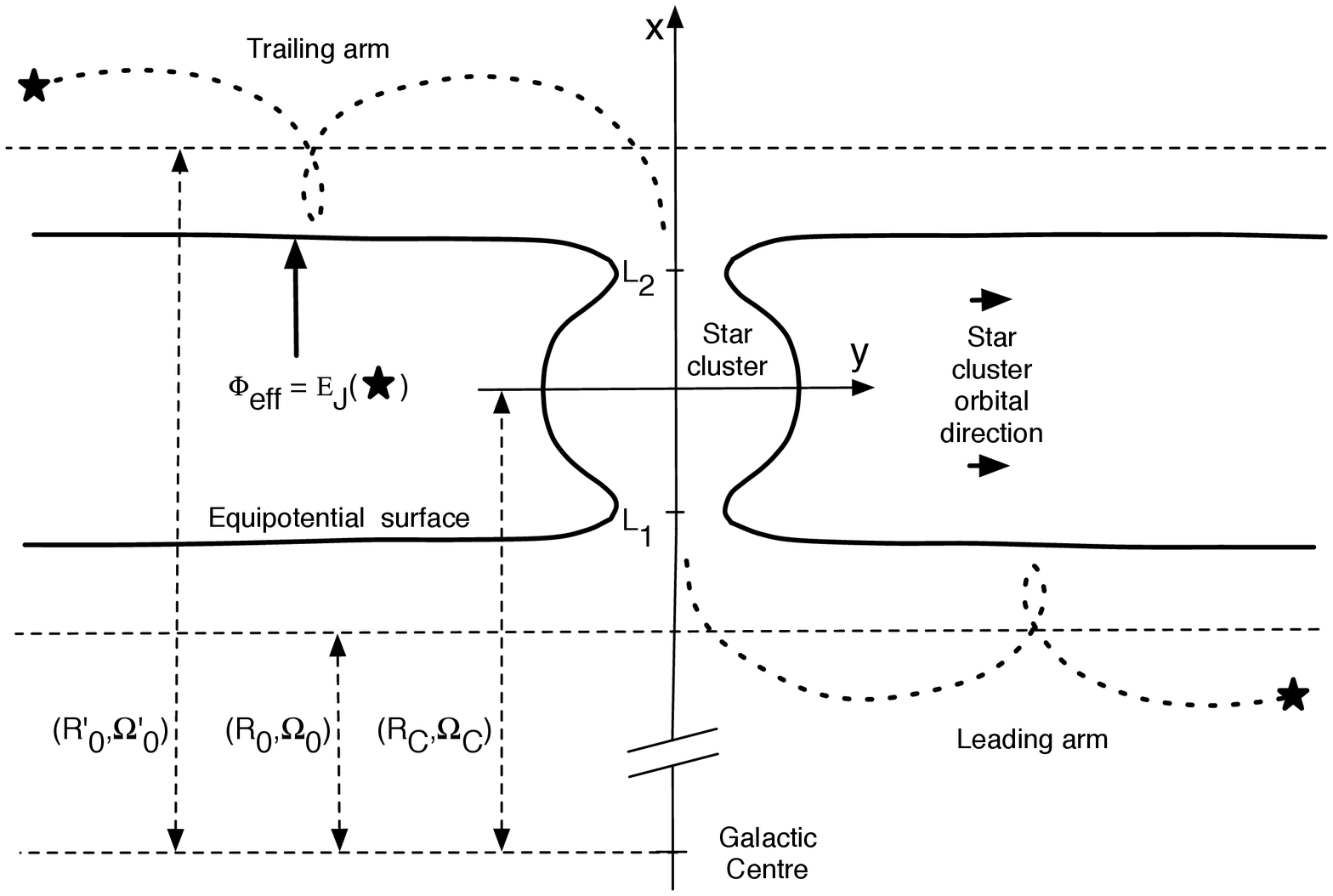}
\caption{Top: Effective potential of a star cluster in the corotating frame. 
L1 and L2 are the Lagrange points.
Bottom: Sketch of escaping stars in the effective potential 
of a star cluster through the Lagrange points L1
and L2.
\label{fig-view}}
%\end{center}
\end{figure} 
%----------------------------------------------------------------------------%

As long as the stars are influenced by the gravitational potential of the
cluster, it is appropriate to use a reference frame corotating with the star
cluster. 
We use cylindrical coordinates $(R,\varphi)$ corotating with the star cluster
and with the origin at the galactic centre. The angular speed is 
$\Omega_\mathrm{C}$ and 
the star cluster centre is at $(R_\mathrm{C},\varphi=0)$. Then we shift the
origin to the star cluster centre and use a local cylindrical coordinate
system with $x=R-R_\mathrm{C}$ and $y=R\varphi$ (for an illustration see figure
~\ref{pos-300}).

In the corotating system we set the zero point of the galactic potential
$\Phi_{g}$ such that the effective potential 
$\Phi_\mathrm{g,eff}(R_\mathrm{C})=\Phi_{g}-\Omega_\mathrm{C}^2 R_\mathrm{C}^2/2=0$ 
vanishes at $R_\mathrm{C}$.
We get 
\bq
\Phi_\mathrm{eff}=\Phi_\mathrm{cl}+\Phi_\mathrm{g} - \Phi_\mathrm{g}(R_\mathrm{C}) -\frac{\Omega_\mathrm{C}^2}{2}(R^2-R_\mathrm{C}^2)
\eq
The Jacobi energy $E_\mathrm{J}$, which is the only known constant
of motion \citep{Bi87}, is given by
\bq
E_\mathrm{J}=E-\Omega_\mathrm{C} L=\Phi_\mathrm{eff}+\frac{v^2}{2}
\eq
with velocity $v$ in the corotating rest frame. The Jacobi energy of the cluster
motion in the galactic potential vanishs.
The effective potential has saddle points at the inner and outer Lagrange points
L1/L2 (Fig. \ref{fig-view}), where the stars are leaking out.
All stars with Jacobi energy  exceeding the critical value 
$E_\mathrm{J,crit}=\Phi_\mathrm{eff}(L1/L2)$ can in principle leave the cluster. 

For the effective potential we get at $y=0$
\bq
\Phi_\mathrm{eff}(x,0)=\frac{\beta_\mathrm{C}^2-4}{2}\Omega_\mathrm{C}^2 x^2+\Phi_\mathrm{cl}(x,0) \label{eq-phieff}
\eq
The tidal radius $r_\mathrm{L}$ is given by the distance of the Lagrange points
to the cluster centre. It is determined
by
\bqn
0&=&\frac{\dd \Phi_\mathrm{eff}}{\dd x}=
	\Omega_\mathrm{C}^2(\beta_\mathrm{C}^2-4)r_\mathrm{L}\pm\frac{GM_\mathrm{cl}(r_\mathrm{L})}{r_\mathrm{L}^2}
\eqn
where $M_\mathrm{cl}(r_\mathrm{L})$ is the cluster mass enclosed in 
$|r_\mathrm{L}|$.
We find the well known equation for the tidal radius
\bq
r_\mathrm{L}^3=\frac{\pm GM_\mathrm{cl}}{(4-\beta_\mathrm{C}^2)\Omega_\mathrm{C}^2} \label{eq-rL}
\eq
where we assumed that the full cluster mass is enclosed in $|r_\mathrm{L}|$.
The effective potential at the Lagrange points is
\bq
\Phi_\mathrm{eff}(|r_\mathrm{L}|,0)=E_\mathrm{J,crit}=-\frac{3}{2}
	(4-\beta_\mathrm{C}^2)\Omega_\mathrm{C}^2 r_\mathrm{L}^2 =
	-\frac{3}{2}\frac{GM_\mathrm{cl}}{|r_\mathrm{L}|} \label{eq-phieffL}
\eq
The last expression shows that the contribution from the star cluster potential
is twice that of the effective potential of the Galaxy.

A star starting near L1 or L2 with velocity $v_\mathrm{L}$ escapes at constant
Jacobi energy but with changing energy and angular momentum
until the cluster potential can be neglected.
Then the position ($x,y$) and velocity $v=(v_\mathrm{r},v_\mathrm{t})$ in the tidal tail
are related to ($r_\mathrm{L},0$) and $v_\mathrm{L}$ by
\bq
E_\mathrm{J}=\Phi_\mathrm{eff}(r_\mathrm{L},0)+\frac{v_\mathrm{L}^2}{2}=\Phi_\mathrm{g,eff}(x)+\frac{v^2}{2}
\eq
leading to
\bq
x^2=3r_\mathrm{L}^2+\frac{\Delta(v^2)}{(4-\beta_\mathrm{C}^2)\Omega_\mathrm{C}^2}
\eq
or 
\bq
\frac{x^2}{r_\mathrm{L}^2}=3+\frac{\Delta(v^2)}{GM_\mathrm{cl}/|r_\mathrm{L}|}
\eq
with $\Delta (v^2) =v^2-v_\mathrm{L}^2$.
Stars moving along the equipotential surface ($\Delta (v^2) =0$) yield
as initial position $x=\sqrt{3}r_\mathrm{L}$ and initial velocity essentially tangential
$v_\mathrm{t}\approx v_\mathrm{L}$. 
This approximation fits well with the radial position of the 
equipotential surface through L1/L2 at large distances from the cluster
in Fig. \ref{fig-view}.
Stars moving radially gain kinetic energy ($\Delta v^2 >0$)
resulting in a larger $x$ and stars starting tangentially loose
kinetic energy ($\Delta v^2 <0$) leading to a smaller $x$. 

For a continuous mass loss until dissolution it is necessary that the 
Jacobi energy of bound stars is lifted above the critical value
$E_\mathrm{J,crit}$, which increases due to the mass loss.
There are two physical effects, which are responsible for a continuous mass loss
of the cluster. 
The first one is triggered by the mass loss of the cluster itself. 
Mass loss on a timescale
large compared to the dynamical time of the cluster leads to an increase of
$E_\mathrm{J}$ of the bound stars by
\bq
\frac{\dd E_\mathrm{J}}{\dd t}=\frac{\delta\Phi_\mathrm{cl}}{\delta t}
	\propto\dot{M}_\mathrm{cl}
\eq
But the critical value $E_\mathrm{J,crit}$ increases more slowly, because the
tidal radius decreases with decreasing mass
\bq
\frac{\dd \Phi_\mathrm{eff}(r_\mathrm{L},0)}{\dd t}
	\propto\dot{M}_\mathrm{cl}^{2/3}
\eq
Initiated by mass loss due to stellar evolution or by a few stars
above $E_\mathrm{J,crit}$ mass loss will continue by stars lifted
above the critical value.

The second process is dynamical evolution of the cluster due to 2-body
encounters. With the relaxation timescale stars are scattered above
$E_\mathrm{J,crit}$ and can leave the cluster. The relative importance of
the two effects depend on the mass, number of stars and the structure of the 
cluster.

%%%%%%%%%%%%%%%%%%%%%%%%%%%%%%%%%%%%%%%%%%%%%%%%%%%%%%%%%%%%%%%%%%%%%%%%%%%%%

\subsection{Dynamic parameters of tidal tail stars}
\label{sec-tid}

Since the orbits are
epicycles perturbed by the acceleration of the cluster, the connection of the
initial position and velocity $(r_\mathrm{L},v_\mathrm{L})$ to $(x,v)$ at a later time, when the cluster
potential can be neglected, is very complicated. Here we are interested in the statistics of initial and final properties of the escaping stars.

For the transition from bound stars to escaped stars, we need to combine the
motion in the frame corotating with the cluster $R_\mathrm{C},\Omega_\mathrm{C}$ and that in the
non-rotating reference frame, where we derived the properties of the epicycles 
around $R_0,\Omega_0$. For measuring the
shape and kinematics of the tidal tails we stay in the corotating rest
frame centered at the cluster. Therefore we transform
the epicyclic motion to the corotating frame with respect to  $R_\mathrm{C},\Omega_\mathrm{C}$.

The radial offset $\Delta R_0=R_0-R_\mathrm{C}$ of the epicentre of a star is 
determined by the angular momentum difference $\Delta L=L - L_\mathrm{C}$ 
(see equation~\ref{eq-RofL}).
Here we need only the first order term of $\Delta R_0$ in $\Delta L$, which is
\bqn
\frac{\Delta R_0}{R_\mathrm{C}} &=&
	\frac{2}{\beta_\mathrm{C}^2}\frac{\Delta L}{L_\mathrm{C} }
	= \frac{2}{\beta_\mathrm{C}^2 R_\mathrm{C}}\left( 2x +\frac{ v_\mathrm{t}}{\Omega_\mathrm{C}}\right)
	\label{eq-dR0}
\eqn
Since the epicycles are counterrotating with respect to the disc rotation, the
relative velocity in the tidal tails is smallest at the pericentres (with
respect to the cluster motion). These are the locations where the clumps occur.
The tangential distance $y_0(T)$ of the pericentres in the corotating
frame are determined by
the shear flow of the epicentre motion. The period is $T=2\pi/\kappa_0$ leading
to
\bqn
y_0(T)&=&\frac{2\pi}{\kappa_0} R_0\left(\Omega_0-\Omega_\mathrm{C} \right)
   =\frac{2\pi}{\beta_\mathrm{C}}\frac{R_\mathrm{C}}{\Omega_\mathrm{C}}\Omega_\mathrm{C}'\Delta R_0 \label{eq-y0-dR0}\\
   &=&\frac{2\pi}{\beta_\mathrm{C}}\frac{\beta_\mathrm{C}^2-4}{2}\Delta R_0 =
   \frac{2\pi}{\beta_\mathrm{C}}\frac{\beta_\mathrm{C}^2-4}{\beta_\mathrm{C}^2}\frac{\Delta L}{L_\mathrm{C}}R_\mathrm{C}
    \label{eq-y0-dl} \\
   \nonumber&=&\frac{4\pi}{\beta_\mathrm{C}}\frac{\beta_\mathrm{C}^2-4}{\beta_\mathrm{C}^2}\left( x +\frac{ v_\mathrm{t}}{2\Omega_\mathrm{C}}\right)
\nonumber
\eqn
For the special case of $x=r_\mathrm{L}$ and $v_\mathrm{t}=0$ we recover the equation given in
\citet{Ku08}.
The spread in $\Delta L$ of the escaping stars leads to a corresponding tangential width of the 
first clump. In the succeeding clumps the tangential spread increases linearly and
quickly smears out the pericentre positions over a whole period of $y_0(T)$.

The radial spread of the tidal tail stars is determined by the combined spread in
$\Delta R_0$ and in the amplitudes $r_\mathrm{m}$. The amplitudes are determined by the
second order terms in the energy excess of the stars with respect to the
epicentre energy $E_0$. Relative to the cluster center energy 
$E_\mathrm{C}=\Phi_\mathrm{g}(R_\mathrm{C})+\Omega_\mathrm{C}^2 R_\mathrm{C}^2/2$
and with the help of equation~\ref{eq-e0} we get for the energy difference of 
the epicentre
\bqn
\Delta E_0&=&E_0 - E_\mathrm{C}=\Omega_\mathrm{C}\Delta L- 
	\frac{4-\beta_\mathrm{C}^2}{2\beta_\mathrm{C}^2}
	\frac{\Omega_\mathrm{C}}{L_\mathrm{C}}\Delta L^2 \label{eq-de0}
\eqn

The amplitude of the epicycle is determined by the energy excess with respect 
to the epicentre energy $E_0$ and can be calculated using the Jacobi energy of
the star 
\bqn
E_\mathrm{J}&=&\Phi_\mathrm{g,eff}(R_\mathrm{C}+x)+\frac{v^2}{2} 
	=-\frac{4-\beta_\mathrm{C}^2}{2}\Omega_\mathrm{C}^2 x^2 + \frac{v^2}{2}
\eqn
 (taking into account the zero-point $\Phi_\mathrm{g,eff}(R_\mathrm{C})=0$)
and equation~\ref{eq-de0} leading to
\bqn
\Delta E&=&E-E_0=E_\mathrm{J}+\Omega_\mathrm{C} \Delta L -\Delta E_0 \\
 &=& E_\mathrm{J} +\Omega_\mathrm{C} L_\mathrm{C}
	\frac{4-\beta_\mathrm{C}^2}{2\beta_\mathrm{C}^2}
	\frac{\Delta L^2}{L_\mathrm{C}^2}\label{eq-dedl}\\
	&=&E_\mathrm{J}
	+\Omega_\mathrm{C}^2\frac{\beta_\mathrm{C}^2}{2}\frac{4-\beta_\mathrm{C}^2}{4}
	\Delta R_0^2  \label{eq-de}
\eqn
Now the epicyclic amplitude $r_\mathrm{m}$ is determined by
\bqn
r_\mathrm{m}^2&=& \frac{2 \Delta E}{\beta_\mathrm{C}^2\Omega_\mathrm{C}^2} 
	=\frac{4-\beta_\mathrm{C}^2}{4}\Delta R_0^2 +
	\frac{2}{\beta_\mathrm{C}^2}\frac{E_\mathrm{J}}{\Omega_\mathrm{C}^2}\\
     \nonumber&=&\frac{2}{\beta_\mathrm{C}^2}\left[
     \frac{4-\beta_\mathrm{C}^2}{2\beta_\mathrm{C}^2}
     R_\mathrm{C}^2\frac{\Delta L^2}{L_\mathrm{C}^2}+
     \frac{E_\mathrm{J}}{\Omega_\mathrm{C}^2}\right]\label{eq-rm}
\eqn
The position of the apo/pericentre with respect to the cluster centre is
given by
\bqn
x_\mathrm{m}&=&R_\mathrm{m}-R_\mathrm{C}=\Delta R_0\pm r_\mathrm{m}
\eqn
In terms of position $x$ and velocity $v_\mathrm{r},v_\mathrm{t}$ in the corotating frame
the apo- and pericentre are given by
\bqn
x_\mathrm{m}&=&\frac{1}{\beta_\mathrm{C}^2}\left( 4x +\frac{2 v_\mathrm{t}}{\Omega_\mathrm{C}}\right)\label{eq-xm}\\
   &&\pm \frac{1 }{\beta_\mathrm{C}^2}
   \sqrt{\left(\left(4-\beta_\mathrm{C}^2\right) x - \frac{2v_\mathrm{t}}{\Omega_\mathrm{C}} \right)^2
      +\beta_\mathrm{C}^2\frac{v_\mathrm{r}^2}{\Omega_\mathrm{C}^2}}\nonumber
\eqn
If the star starts at peri/apo-centre (with no radial velocity $v_\mathrm{r}=0$) 
we find the corresponding apo/peri-centre by
\bqn
v_\mathrm{r}=0: & x_\mathrm{m}= \frac{1}{\beta_\mathrm{C}^2}\left(4x+\frac{2 v_\mathrm{t}}{\Omega_\mathrm{C}}\pm
	\left|(4-\beta_\mathrm{C}^2)x-\frac{2 v_\mathrm{t}}{\Omega_\mathrm{C}}\right|\right)
\label{eq:kuepperformula}
\eqn
In \citet{Ku08} the special case of $v_\mathrm{r}=v_\mathrm{t}=0$ and $x=r_\mathrm{L}$ was adopted, where 
the star has a Jacobi energy of 
$E_\mathrm{J}=-GM_\mathrm{cl}/(2r_\mathrm{L})=E_\mathrm{J,crit}/3$.

%%%%%%%%%%%%%%%%%%%%%%%%%%%%%%%%%%%%%%%%%%%%%%%%%%%%%%%%%%%%%%%%%%%%%%%%%%%%%

\section{Numerical modeling}
\label{sec-num}

For the main investigation we used the direct $\phi$GRAPE $N$-body code to calculate the
evolution of star clusters in an analytic Galaxy model. It is described in this
section. For the case of a star cluster near the Galactic centre and for
testing the reliability of the simulations we used a variant of the direct $N$-body6++
code, which is described in Section~\ref{sec-gc}.

%%%%%%%%%%%%%%%%%%%%%%%%%%%%%%%%%%%%%%%%%%%%%%%%%%%%%%%%%%%%%%%%%%%%%%%%%%%%%

\subsection{$\phi$GRAPE $N$-body code}

For the high resolution direct $N$-body simulations at large
galactocentric distances we used the 
specially developed {\bf $\phi$GRAPE} code. The code itself and 
also the special GRAPE hardware is described in more detail in
\citet{Hetal2007}. Here we mention briefly the most important 
special features of the code. The program was already well 
tested with different $N$-body applications including the high 
resolution study of the dynamical evolution of the galactic centre 
with a binary (or single) Super-massive Black Hole 
\citep{BMS2005, BMSB2006, MBL2007}. 
The same code was also recently used to study the shape parameters of 
a large set of rotating open star clusters 
\citep{Khar08}\footnote{The present 
version of the code will be publicly available 
from one of the authors FTP site: 
{\tt ftp://ftp.ari.uni-heidelberg.de/staff/berczik/ 
phi-GRAPE-cluster/code-paper/}.}. 

The program acronym {\bf $\phi$GRAPE} means: {\bf P}{\it arallel} 
{\bf H}{\it ermite} {\bf I}{\it ntegration} with {\bf GRAPE}. The 
serial and parallel version of the program has been written from scratch 
in ANSI-C and 
uses the standard MPI library for communication. For the 
calculation of the star cluster dynamics in the 
galactic potential we use the parallel GRAPE systems 
built at the Astronomisches Rechen-Institut in 
Heidelberg\footnote{GRACE: {\tt http://www.ari.uni-heidelberg.de/grace}}, 
and at the Main Astronomical Observatory in Kiev\footnote{GOLOWOOD: 
{\tt http://www.mao.kiev.ua/golowood/eng}}. 

The program uses the 4-th order Hermite integration scheme for 
the particles with hierarchical individual block timesteps, 
together with the parallel usage of GRAPE6a cards for the 
hardware calculation of the acceleration ${\bf  a}$ and the 
first time derivative of the acceleration ${\bf \dot{{a}}}$ 
(this term is usually called 'jerk' in the $N$-body community).

For the simulation of star clusters in the tidal field of the 
Galaxy an analytic external potential is added. We use an axi-symmetric three 
component model, where bulge, disc and halo are described by 
Plummer-Kuzmin models \citep{MN1975} with
\begin{equation}
  \Phi(R,z) = - \frac{ G \cdot M }{ \sqrt{R^2 + (a + \sqrt{b^2 + z^2} )^2} },
  \label{eq-gal}
\end{equation}
where $b$ is a measure of the core radius and $a$ a measure of the flattening.
%----------------------------------------------------------------------------%
\begin{table}
\caption{The list of galaxy component parameters. The first column gives the
component, the second the mass, and the third and fourth the
Plummer-Kuzmin parameters (equation~\ref{eq-gal}).}
\label{gal-par}
\begin{center}
\begin{tabular}{lcrr}
\hline\noalign{\smallskip}
 Mass component & M [M$_\odot$] & $a~[{\rm kpc}]$ & $b~[{\rm kpc}]$ \\
\noalign{\smallskip}
\hline
\noalign{\smallskip}
 Bulge & $1.4 \cdot 10^{10}$ & 0.0 &  0.3 \\
 Disk  & $9.0 \cdot 10^{10}$ & 3.3 &  0.3 \\
 Halo  & $7.0 \cdot 10^{11}$ & 0.0 & 25.0 \\ 
\hline
\end{tabular}
\end{center}
\end{table}
%----------------------------------------------------------------------------%

\begin{figure}
  \begin{center}
% \vspace{-0.2\textwidth}
  \includegraphics[width=0.45\textwidth]{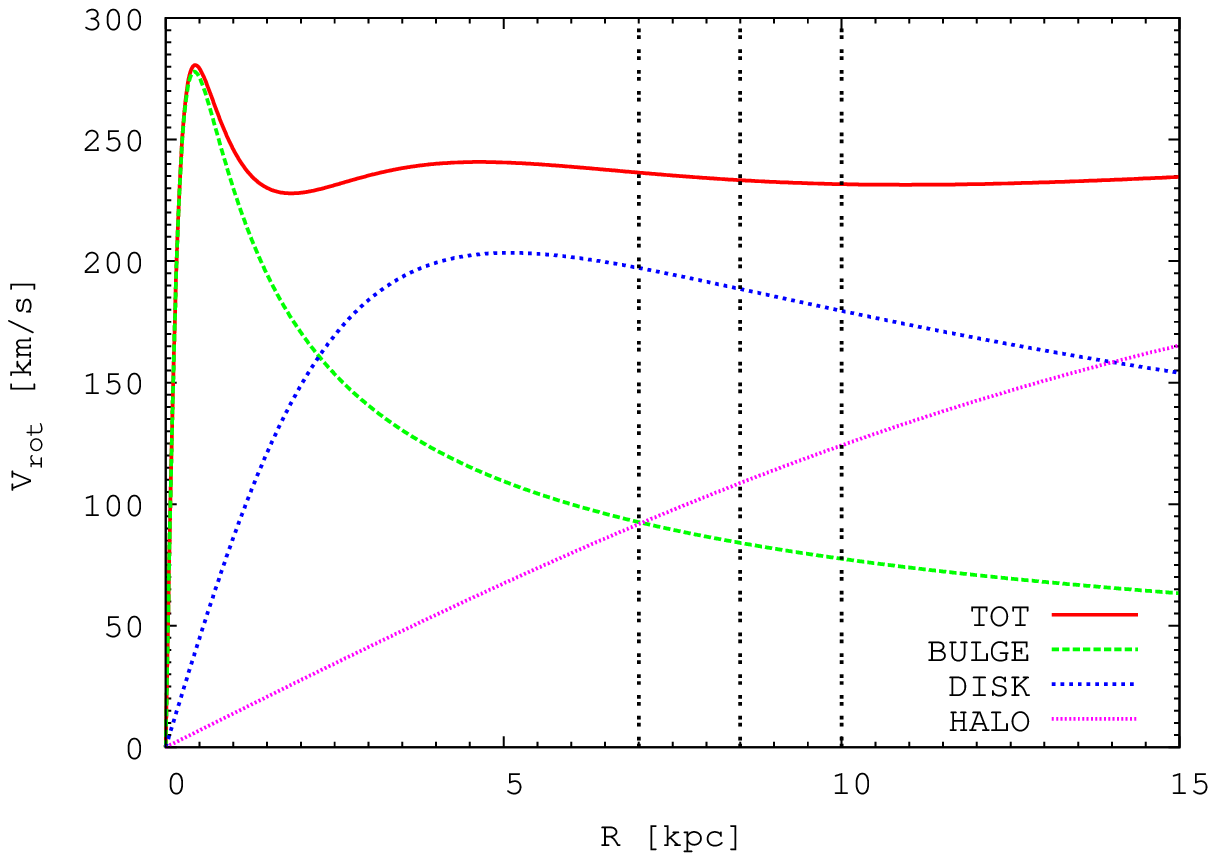}
  \includegraphics[width=0.45\textwidth]{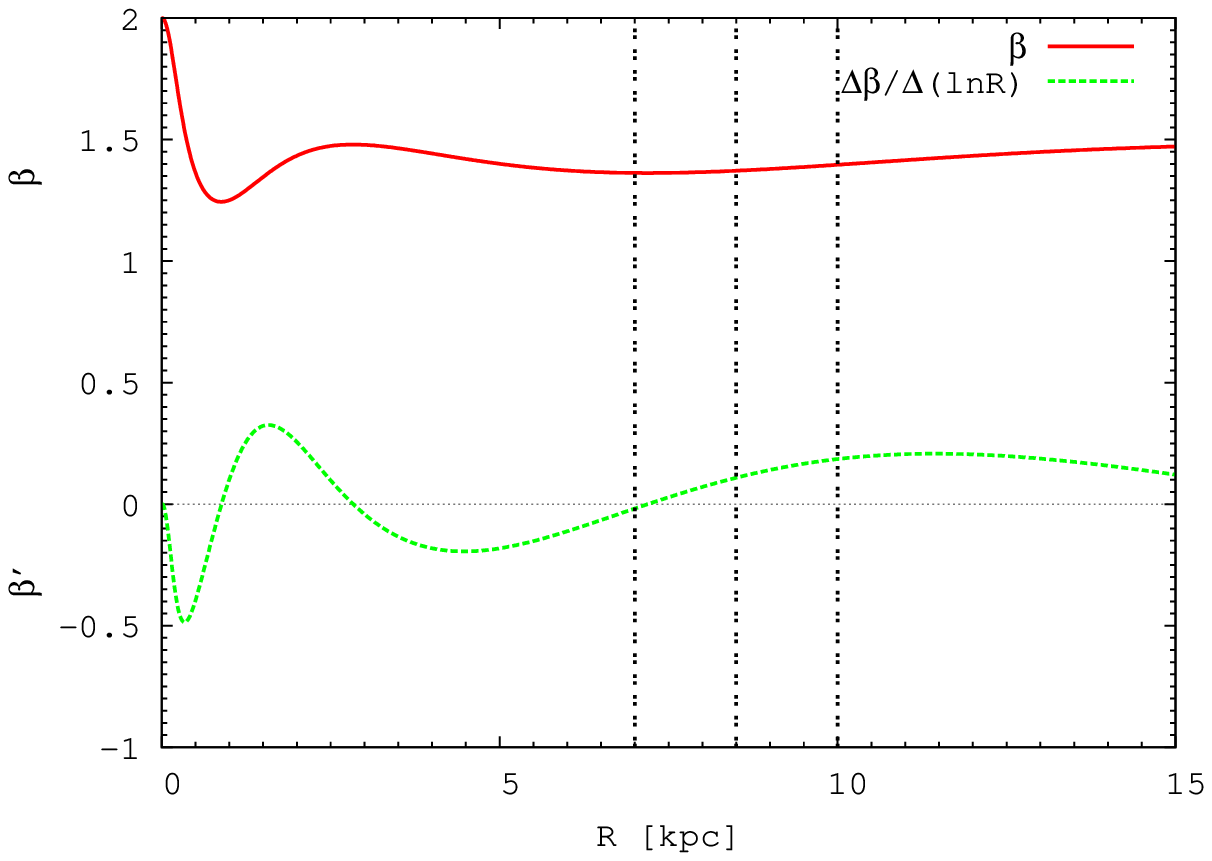}
  \end{center}
  \caption{Top: Rotation curve of the Galaxy model and the contributions from bulge, 
  disk and halo. Bottom: Epicyclic frequency parameter $\beta$ and  the logarithmic 
  derivative $\beta'$. The vertical dashed lines mark the orbits of the star
  clusters selected for the investigation.
  \label{fig-rot}}
\end{figure}

For the parameters we use similar values as in \citet{DC1995} 
with slightly corrected masses (see Table \ref{gal-par}) to 
reproduce the observed Milky Way rotation curve in the 
solar neighbourhood. The rotation curve, the epicyclic frequency
 $\beta=\kappa/\Omega$ (normalized to the orbital frequency $\Omega$)
and the logarithmic derivative $\beta'=\dd \beta/\dd\ln R$ are shown in Fig. 
\ref{fig-rot}.

%----------------------------------------------------------------------------%
%----------------------------------------------------------------------------%

%%%%%%%%%%%%%%%%%%%%%%%%%%%%%%%%%%%%%%%%%%%%%%%%%%%%%%%%%%%%%%%%%%%%%%%%%%%%%

\subsection{Initial conditions for the star cluster}

The star clusters are modeled star by star using a Salpeter 
IMF \citep{S1955} in the mass range of $0.08\le m/\msun\le 8 $. 
We include a simple model for stellar evolution from 
\citet{vdHG1997} and distribute the stellar mass loss uniformly over the 
metallicity dependent stellar lifetimes \citep{RVN1996}.

For the generation of the initial particle distribution and 
velocities we use a nonrotating King model with W$_0=6.0$. 
 In the $N$-body code the
physical quantities are normalized by $G=M_\mathrm{cl}=1$ and total energy
$E^{TOT}=-0.25$ as first introduced by \citet{AHW1974}.

After generating the dimensionless parameters for the 
cluster data we set the different physical mass \& half-mass 
radius parameters for the clusters. In general these two 
parameters can be set independently, but in order to reduce the number of 
free parameters we decided to use a physically 
motivated relation between initial mass and initial half-mass 
radius of the star clusters. For this purpose we use the 
extension of the well known mass vs. radius relation observed in 
molecular clouds and clumps in our Milky Way. For relevant 
references see the list of observational and theoretical papers 
\citet{L1981, SRBY1987, M1990, TH1993, IK2000}. Using such an 
approximation we can write down a scaling relation for our 
initial cluster size and mass. For fixing the size of the cluster 
we use the radius $r_\mathrm{k}$ containing 60\% of the cluster 
mass (because this radius is approximately independent of 
rotation for future extension to rotating King models). We set 
\begin{equation}
 r_\mathrm{k} \approx 100 \cdot \sqrt{ \frac{ M_\mathrm{cl} }{ 10^6 \msun } } 
 ~[{\rm pc}]
\end{equation}

For the used three physical masses M$_\mathrm{cl}$ = 10$^3$, 
5$\cdot$10$^3$ and 10$^4$ [M$_\odot]$ we get the corresponding 
radii $r_\mathrm{k}$ = 3.0, 7.0 and 10 [pc]. We do not adapt the 'tidal 
radius' of the King models, where the density vanishes, to the 
tidal radius $r_\mathrm{L}$ of the galactic field. 
The starting point of the star cluster orbit is determined by 
the position in the galactic disk (0.0, $R_\mathrm{C}$, 0.0) 
with corresponding velocity (-$V_\mathrm{C}$, 0.0, 0.0) 
added to each star. 
The star cluster is nonrotating and has 
exactly the angular momentum $V_\mathrm{C}R^2_\mathrm{C}$. If $V_\mathrm{C}$ is
exactly the circular velocity at $R_\mathrm{C}$, then the angular momentum
corresponds to $L_\mathrm{C}$ of the circular orbit at 
$R_\mathrm{C}$. In contrast \citet{Fu00} started with clusters in the 
corotating frame leading to an additional spin of the star cluster. 
For practical reasons the initial velocity was slightly smaller then the correponding
circular speed at $R_\mathrm{C}$ by neglecting the decimal places. Therefore the
star clusters in models 01--09 started at apo-centre and moved on an epicycle
relative to the cluster epicentre motion. It turned out that the
galactocentric distance variations of the cluster orbits are comparable to the 
tidal radii of the star clusters. Since the Jacobi energy is conserved only in a
rest frame with constant angular speed, the origin of the coordinate system must
be determined by the epicentre motion of the cluster and therefore the cluster
centre is moving on an
epicycle in that coordinate system. The main effect on the evolution of the
tidal tails is an additional periodic force with the epicyclic frequency.
In order to test the effect of this 'resonant forcing' we set up model 10 which
is on an exact circular orbit.
The only difference to model 08 is the larger initial velocity by 
0.3\,km/s. The differences between these two models in the mass loss rate
and in the position and strength of the tidal clumps are negligible.
Therefore we use model 10 as the fiducial model for the detailed investigations. The other models are
used to investigate the parameter dependences of the tidal tail 
structure. 
The cluster parameters of the models are 
listed in Table~\ref{model-list}. 

%----------------------------------------------------------------------------%
\begin{table*}
\caption{The model parameters of all runs. Column 1 gives the number
of the model, columns 2--4 are initial mass, number of particles and scale radius
of the cluster, columns 5 and 6 are the initial position and velocity of the
cluster, column 7 is the epicycle parameter at distance $R_\mathrm{C}$, column
 8 is the initial Lagrange radius of the cluster.}
\label{model-list}
%\begin{flushleft}
\begin{center}
\begin{tabular}{crrrrccc}
\hline\noalign{\smallskip}
 \# & M$_\mathrm{cl}$ [M$_\odot$] & N & r$_{k}$ [pc] & R$_\mathrm{C}$ [kpc] & V$_\mathrm{C}$ [km/s] & $\beta$ & r$_\mathrm{L}$ [pc] \\
\noalign{\smallskip}
\hline
\noalign{\smallskip}
 01 & 10$^3$         & 4040   &  3.0 &  7.0 & 236     & 1.363 & 12.08 \\
 02 & 10$^3$         & 4040   &  3.0 &  8.5 & 233     & 1.372 & 13.93 \\
 03 & 10$^3$         & 4040   &  3.0 & 10.0 & 231     & 1.396 & 15.78 \\
\hline 
 04 & 5$\cdot$10$^3$ & 20202  &  7.0 &  7.0 & 236     & 1.363 & 20.66 \\
 05 & 5$\cdot$10$^3$ & 20202  &  7.0 &  8.5 & 233     & 1.372 & 23.82 \\
 06 & 5$\cdot$10$^3$ & 20202  &  7.0 & 10.0 & 231     & 1.396 & 26.98 \\
\hline
 07 & 10$^4$         & 40404  & 10.0 &  7.0 & 236     & 1.363 & 26.04 \\
 08 & 10$^4$         & 40404  & 10.0 &  8.5 & 233     & 1.372 & 30.00 \\
 09 & 10$^4$         & 40404  & 10.0 & 10.0 & 231     & 1.396 & 34.00 \\
\hline
 10 & 10$^4$         & 40404  & 10.0 &  8.5 & 233.297 & 1.372 & 30.00 \\
\hline
\end{tabular}
%\end{flushleft}
\end{center}
\end{table*}
%----------------------------------------------------------------------------%

%%%%%%%%%%%%%%%%%%%%%%%%%%%%%%%%%%%%%%%%%%%%%%%%%%%%%%%%%%%%%%%%%%%%%%%%%%%%%

\section{Results}
\label{sec-res}

We discuss in detail the properties of model 10 on an exact circular
orbit. In this case the origin of the corotating coordinate system is at the
cluster centre. 
Firstly we analyze the mass loss and the Jacobi energy distribution 
of the bound stars. Then we determine  Jacobi energy, angular 
momentum and energy excess of the stars in the tidal tails and 
compare the predictions with the clump positions and widths. 
Then we discuss the parameter dependence of the tidal tail structure.
Finally we present a numerical simulation of a star cluster near the galactic
centre to demonstrate the generality of the theory.

%%%%%%%%%%%%%%%%%%%%%%%%%%%%%%%%%%%%%%%%%%%%%%%%%%%%%%%%%%%%%%%%%%%%%%%%%%%%%

\subsection{Cluster mass loss}
\label{sec-mdot}

\begin{figure}
  \begin{center}
% \vspace{-0.2\textwidth}
  \includegraphics[width=0.45\textwidth]{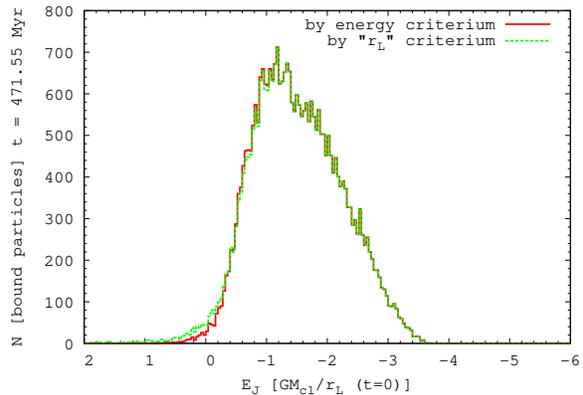}
  \end{center}
  \caption{Jacobi energy distribution of dynamical
cluster members and of stars inside a sphere with radius $|r_\mathrm{L}|$
 at t=471\,Myr of model 10. Note that lower values of $E_\mathrm{J}$ are to the
 right.}
  \label{distr-diff}
\end{figure}

There is no unique definition of bound stars for star clusters in 
tidal fields. The main reason is that many stars with Jacobi 
energy exceeding the critical value remain for a long time in the 
vicinity of the cluster. For determining the mass loss rate and 
for visualisation we use a rather conservative measurement by 
using an energy criterion in the comoving, but not corotating, 
reference system\footnote{The video snapshots 
from all the simulations will be publicly 
available from the FTP site: 
{\tt ftp://ftp.ari.uni-heidelberg.de/staff/berczik/
phi-GRAPE-cluster/video-paper/pos/}.}. 
We assume, that all particles which have a 
negative relative energy in the cluster potential are still bound 
to the cluster: 
$$
  |E^\mathrm{GRA}_\mathrm{i}|_\mathrm{cl} > E^\mathrm{KIN}_\mathrm{i}. 
$$  
An inspection of the particle distribution shows that this 
criterion coincides approximately with stars inside $|r_\mathrm{L}|$. On 
the basis of this criterium we create the list of particles denoted by
'dynamical' cluster members.
In Fig.~\ref{distr-diff} the difference 
in the Jacobi energy distribution of dynamical
cluster members and of stars inside a sphere with radius $|r_\mathrm{L}|$ are
shown. 
There are only small deviations at the high energy end.

%----------------------------------------------------------------------------%
%----------------------------------------------------------------------------%

%----------------------------------------------------------------------------%
\begin{figure}
  \begin{center}
% \vspace{-0.2\textwidth}
  \includegraphics[width=0.45\textwidth]{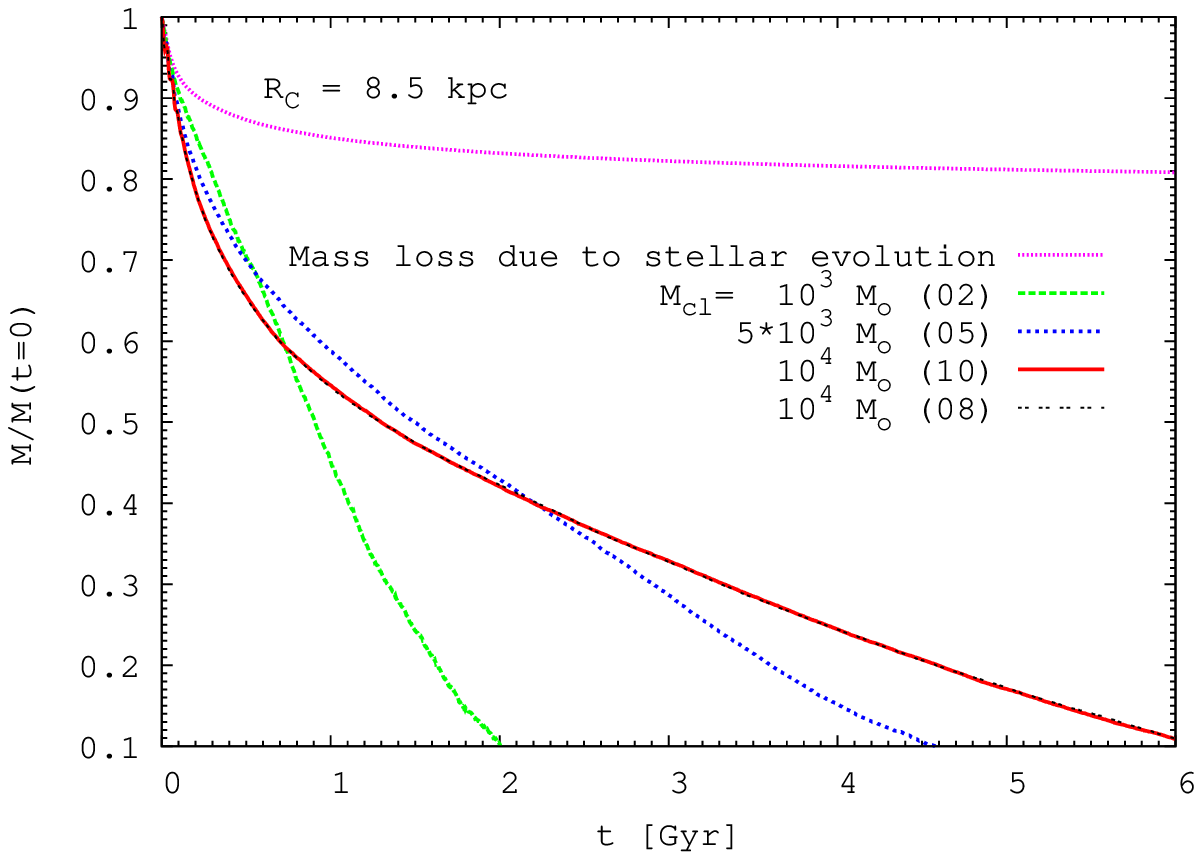}
  \includegraphics[width=0.45\textwidth]{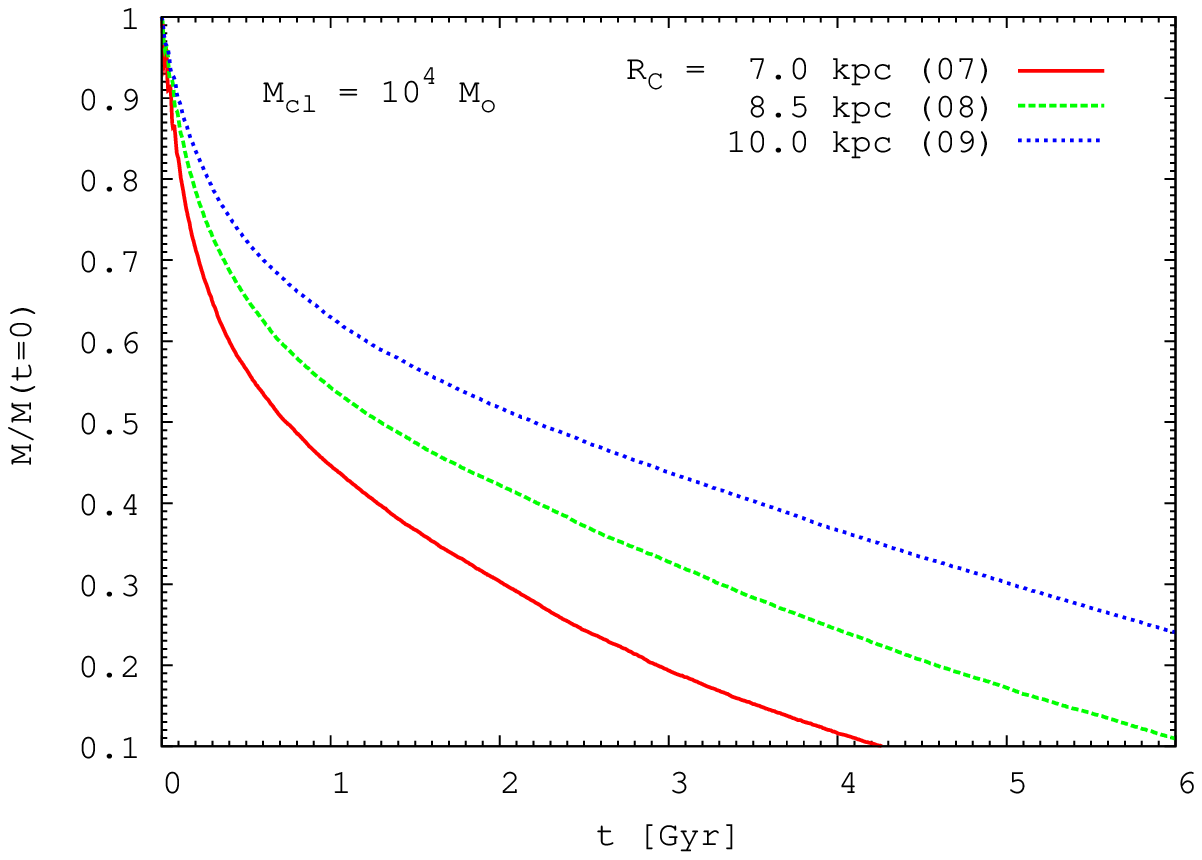}
  \end{center}
  \caption{The time evolution of bound mass of the different models.
  Top: All models at distance $R_\mathrm{C}=8.5$\,kpc. The top-most line shows the
  mass loss due to stellar evolution of all stars.
  Bottom: Models with initial mass $M_\mathrm{cl}=10^4\,\msun$ at different 
  distances to the galactic centre.}
  \label{mass-loss}
\end{figure}
%----------------------------------------------------------------------------%

In Figure~\ref{mass-loss} we present the mass evolution of the star clusters. 
In the upper panel the models at $R_\mathrm{C}=8.5$\,kpc are shown. Mass loss of model
08 at a slightly eccentric orbit shows a very small modulation compared to the 
corresponding model 10 on the exact circular orbit. The top line shows the
mass loss of all stars due to stellar evolution for model 10.
The bottom panel shows the mass evolution of models with initial mass
$M_\mathrm{cl}=10^4\msun$ at different galactocentric distances.

%----------------------------------------------------------------------------%
\begin{figure}
  \begin{center}
% \vspace{-0.2\textwidth}
  \includegraphics[width=0.23\textwidth]{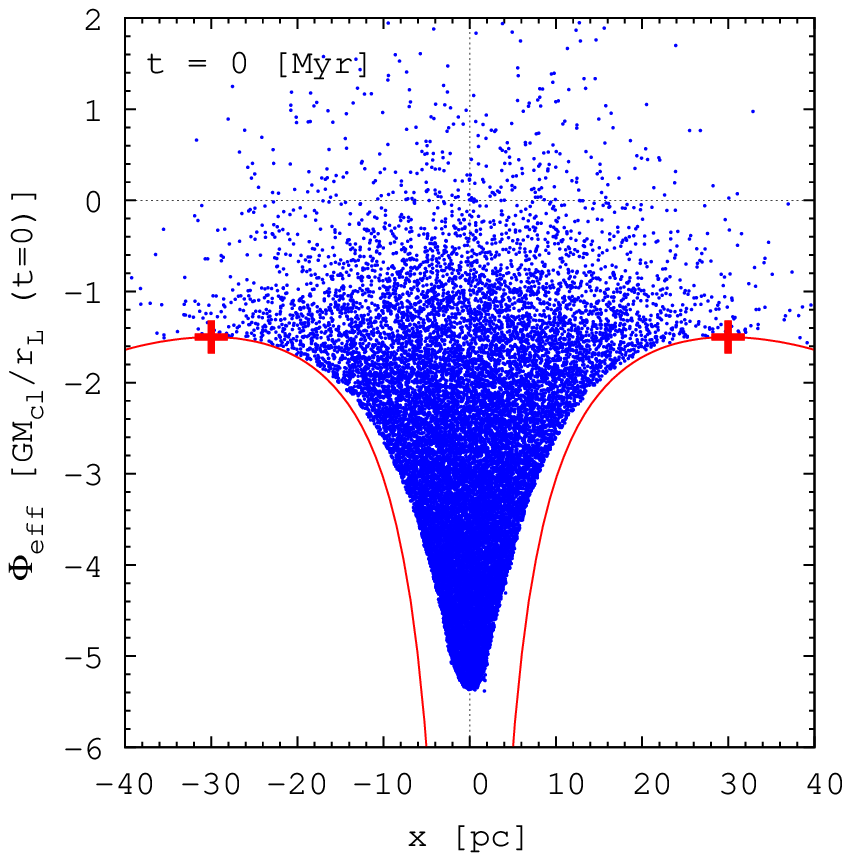}
  \includegraphics[width=0.23\textwidth]{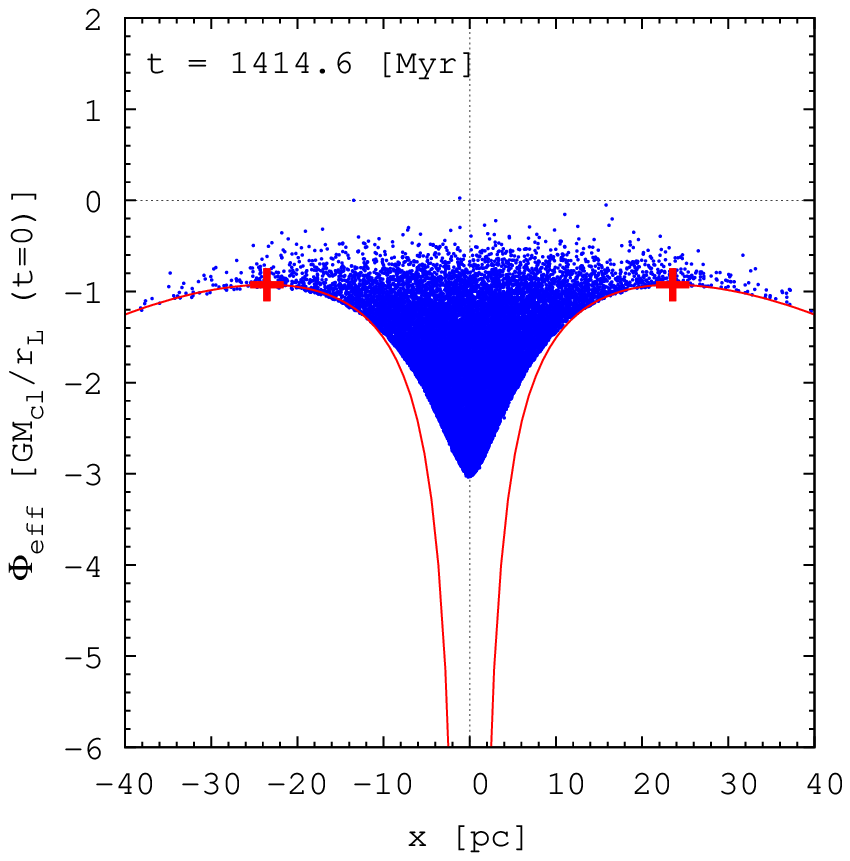}
  \includegraphics[width=0.23\textwidth]{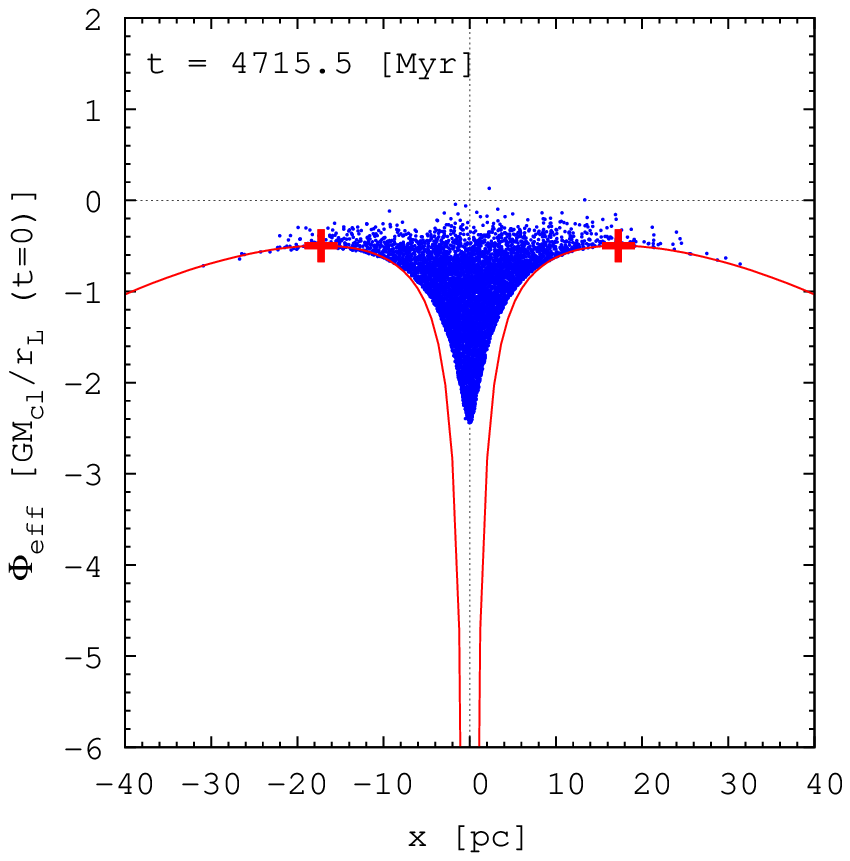}
  \includegraphics[width=0.23\textwidth]{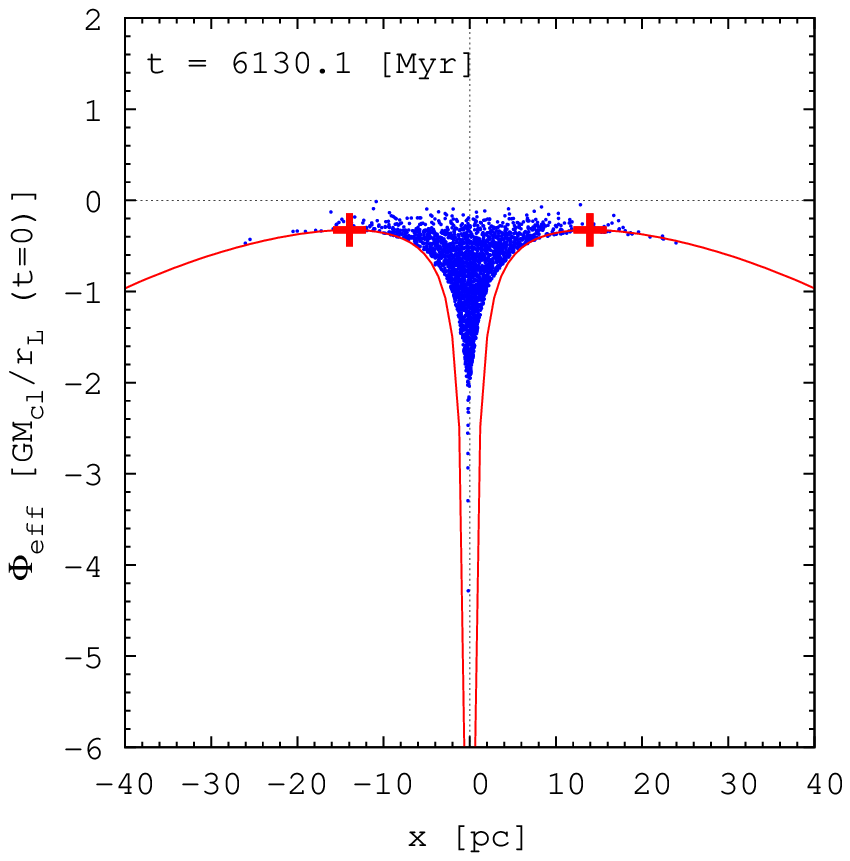}
  \end{center}
  \caption{The effective potential of all bound particles for different times of model 10. 
  The full (red) lines show the analytic approximation according to 
  equation~\ref{eq-phieff}. 
  The vertical tics show the position and energy of the Lagrange points $r_\mathrm{L}$.}
  \label{eff-pot-evol}
\end{figure}
%----------------------------------------------------------------------------%

Figure~\ref{eff-pot-evol} shows the effective potential of bound stars
in units of initial $GM_\mathrm{cl}/r_\mathrm{L}$ at different times as a 
function of radial position $x$ covering a mass loss range of 50\% to 90\%. 
At the last
timestep the cluster mass is already reduced to 10\% of the initial mass.
The full lines show the approximation of $\Phi_\mathrm{eff}(x,0,0)$ 
from equation~\ref{eq-phieff} 
using a point mass potential for the star cluster. It is a lower boundary of 
$\Phi_\mathrm{eff}(x,y,z)$ in projection and shows a perfect agreement in the 
vicinity of the tidal radius $r_\mathrm{L}$, which is marked by the tics. 
According to equations~\ref{eq-rL} and \ref{eq-phieffL} 
the tidal radius $r_\mathrm{L}$ and the
critical energy $E_\mathrm{J,crit}$ scale with $M_\mathrm{cl}^{1/3}(t)$
and $M_\mathrm{cl}^{2/3}(t)$, respectively. 

\begin{figure}
  \begin{center}
% \vspace{-0.2\textwidth}
  \includegraphics[width=0.23\textwidth]{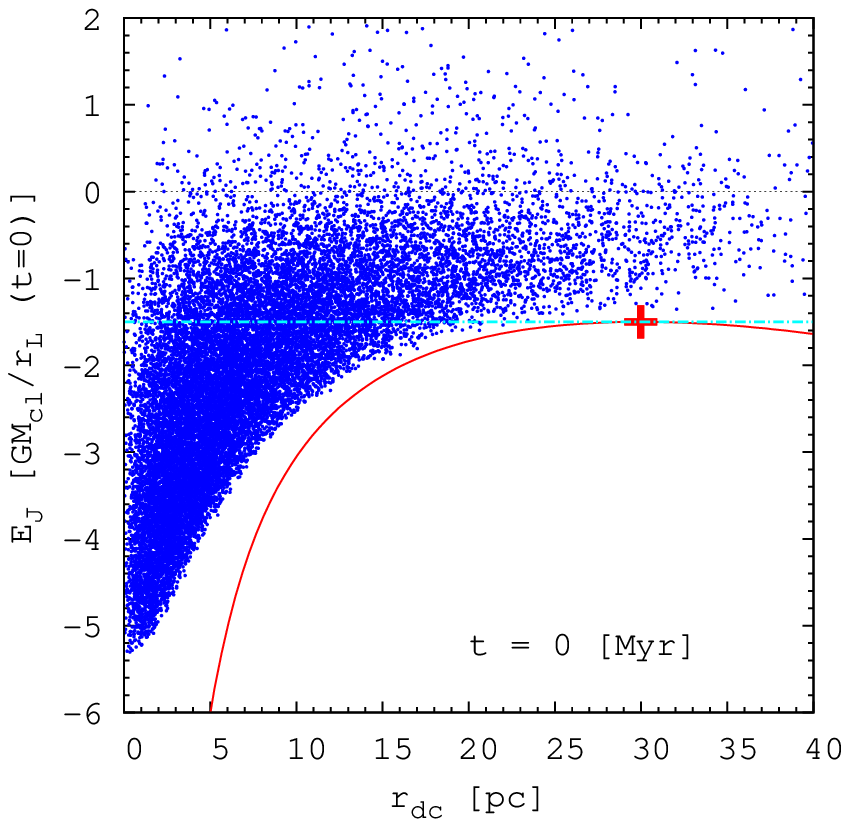}
  \includegraphics[width=0.23\textwidth]{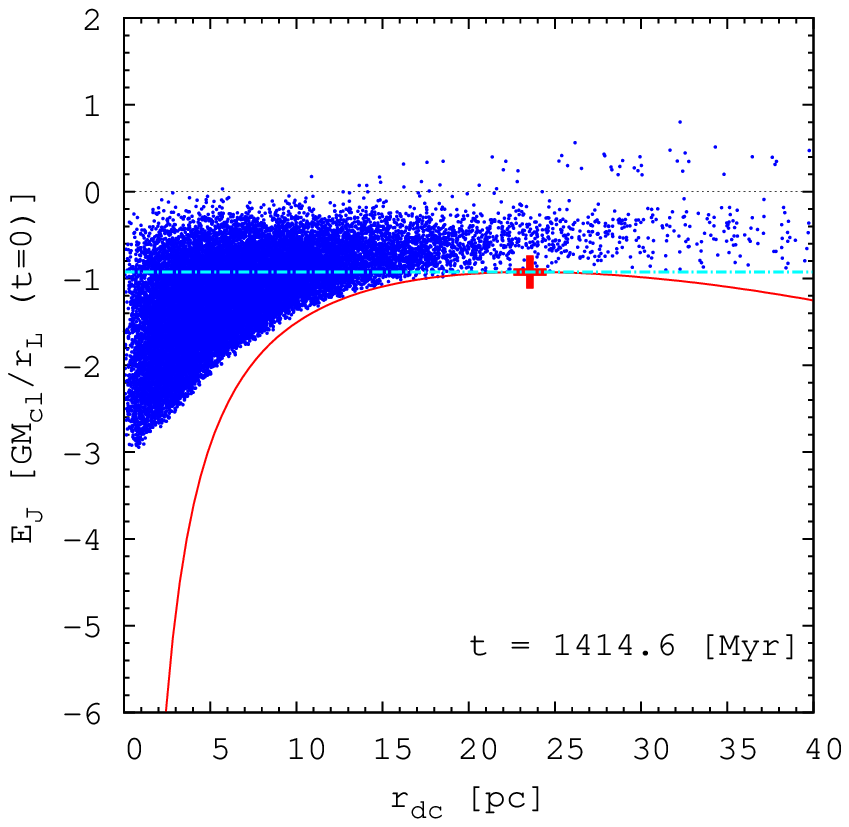}
  \includegraphics[width=0.23\textwidth]{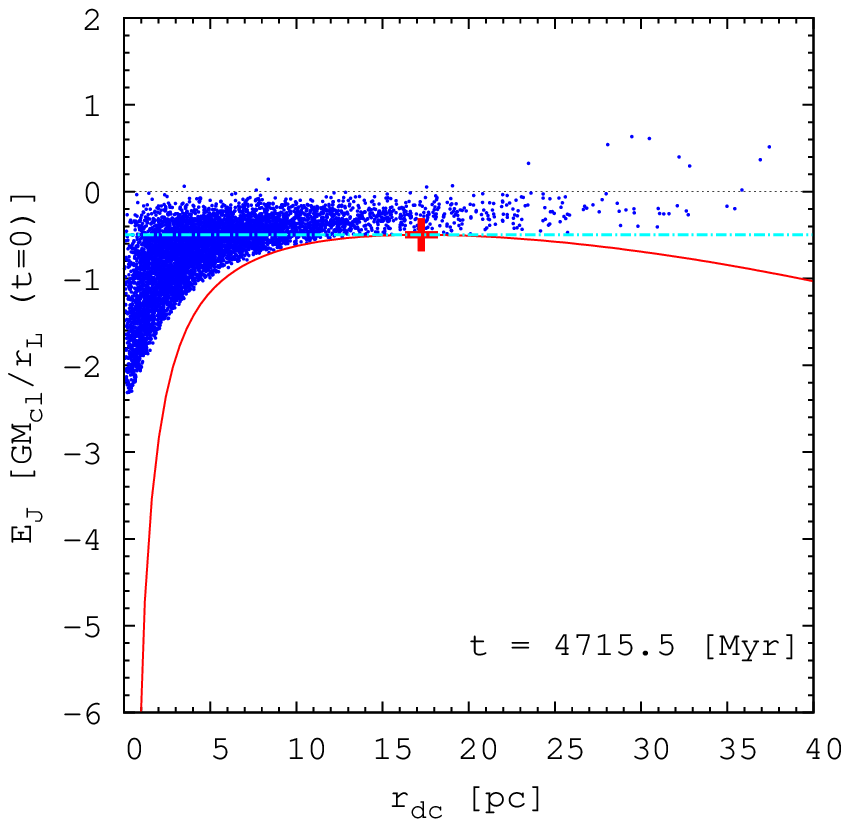}
  \includegraphics[width=0.23\textwidth]{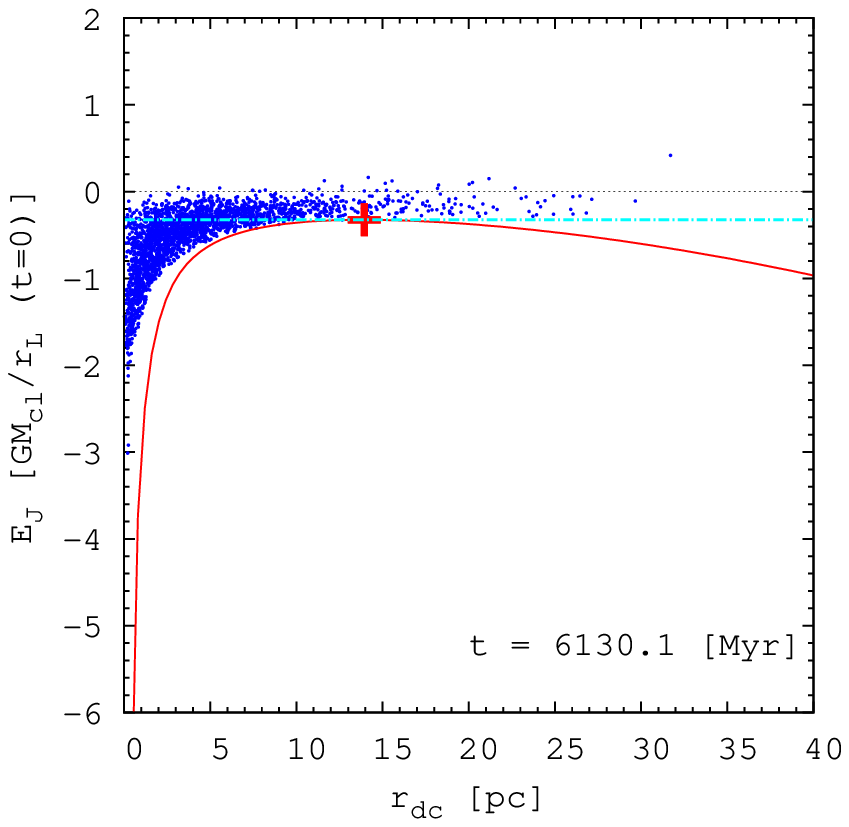}
  \end{center}
  \caption{The $E_\mathrm{J}$ distribution of all bound stars as function of 
  distance to
  the cluster centre $r_\mathrm{dc}$ defined by the density centre for 
  different times of model 10. The full (red) lines are the same
  as in Figure~\ref{eff-pot-evol}. The dashed (cyan) lines mark the critical
  Jacobi energy.}
  \label{ejac-evol}
\end{figure}

\begin{figure}
  \begin{center}
% \vspace{-0.2\textwidth}
  \includegraphics[width=0.45\textwidth]{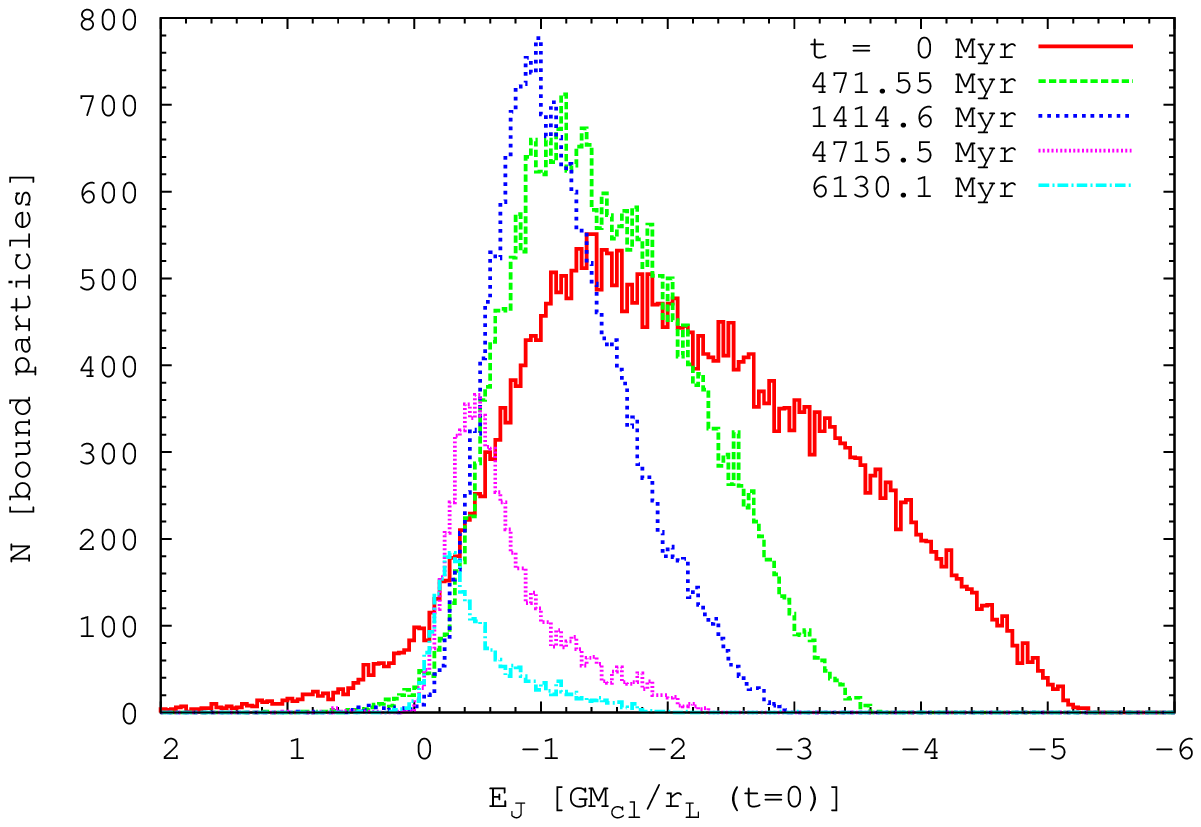}
  \includegraphics[width=0.45\textwidth]{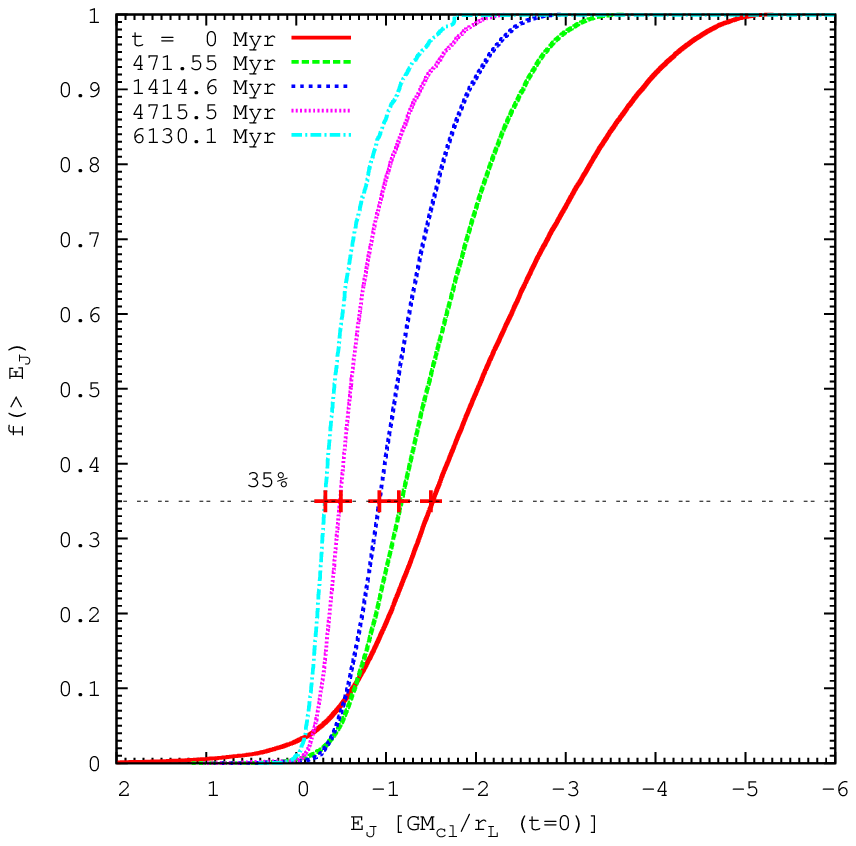}
  \end{center}
  \caption{Top: The Jacobi energy distribution of bound particles at different 
  times of model 10.
  Bottom: The cumulative distribution of bound particles as function of
  decreasing Jacobi energy for different times. The critical 
  Jacobi energy is marked by crosses.
  }
  \label{distr-selected}
\end{figure}

Figure~\ref{ejac-evol} shows the Jacobi energy $E_\mathrm{J}$ in 
units of initial $GM_\mathrm{cl}/r_\mathrm{L}$ of all bound stars 
as in Figure~\ref{eff-pot-evol} but as function of distance to the cluster
centre $r_\mathrm{dc}$. The full lines are the same functions as in
Figure~\ref{ejac-evol}. At all times there is a considerable number of stars
exceeding the critical value $E_\mathrm{J,crit}$ which is marked by the dashed
horizontal lines. These potential escapers are well distributed all over the
cluster.
In Figure~\ref{distr-selected} we quantify the energy distribution of the bound
stars.
The histograms of $E_\mathrm{J}$ in the upper panel of
Figure~\ref{distr-selected} show the temporal evolution
 of bound stars, which demonstrates that the maximum of the distribution is near
 the critical energy with a large fraction of stars above the critical energy.
The lower panel shows the
cumulative distributions of $E_\mathrm{J}$ starting at the high energy end. 
The critical values are marked by crosses showing that at all times up to the 
dissolution the fraction of 'potential escapers' is 35\%. 
In order to measure the importance of 2-body encounters on
the evolution of $E_\mathrm{J}$ more detailed star-by-star investigations 
are necessary. This is beyond the scope of this paper.

%----------------------------------------------------------------------------%
%----------------------------------------------------------------------------%

%----------------------------------------------------------------------------%
%----------------------------------------------------------------------------%

%%%%%%%%%%%%%%%%%%%%%%%%%%%%%%%%%%%%%%%%%%%%%%%%%%%%%%%%%%%%%%%%%%%%%%%%%%%%%

\subsection{Tidal tail clumps}
\label{sec-clump}

\begin{figure*}
  \begin{center}
% \vspace{-0.2\textwidth}
  \includegraphics[width=1.1\textwidth]{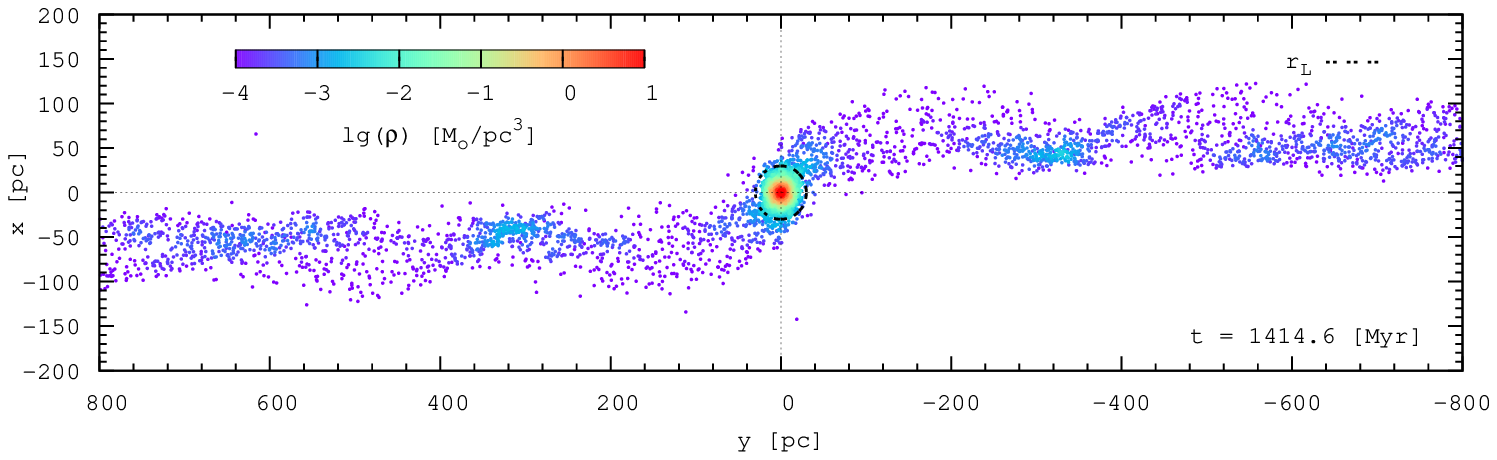}
  \includegraphics[width=1.1\textwidth]{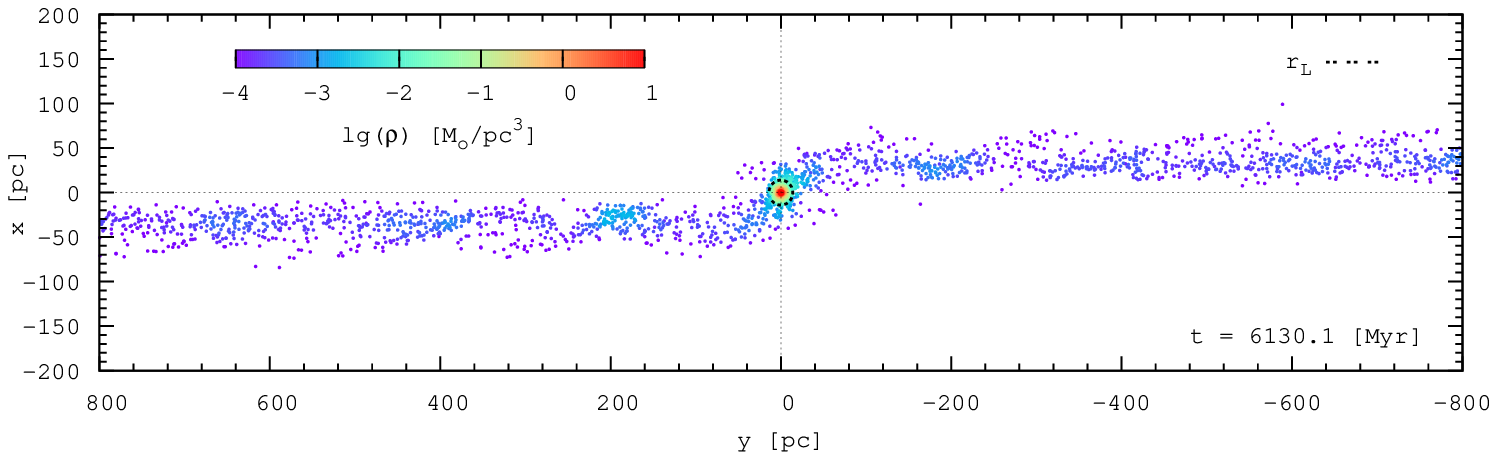}
  \end{center}
  \caption{Positions and local densities of the stars of model 10 at an early 
  and late time with about 50\% and 10\% bound mass, respectively.}
  \label{pos-den}
\end{figure*}

The structure of the tidal tails are determined by the angular
momentum offset $\Delta L$ and energy excess $\Delta E$ of the stars. 
All stars move on
epicycles with angular frequency $\kappa$. The epicentres are determined by
$\Delta L$ (see equation~\ref{eq-dR0}) and the amplitudes by $\Delta E$
(see equation~\ref{eq-rm}). Clumps form at the pericentres, where the streaming
velocity is minimal.
Figure~\ref{pos-den} shows the clumps in the tidal tails for two different
times. The local density at the position of each star is determined by the 
neighbour criterion of \citet{CH1985} using 10 neighbours. It is 
colour coded in logarithmic scale. Note that the clump density does not decrease with
decreasing cluster mass. The circle marks the tidal radius. The structure is
similar in the trailing and leading arm as expected from the symmetry in the 
epicyclic approximation.

%----------------------------------------------------------------------------%
%----------------------------------------------------------------------------%

\begin{figure}
  \begin{center}
% \vspace{-0.2\textwidth}
  \includegraphics[width=0.45\textwidth]{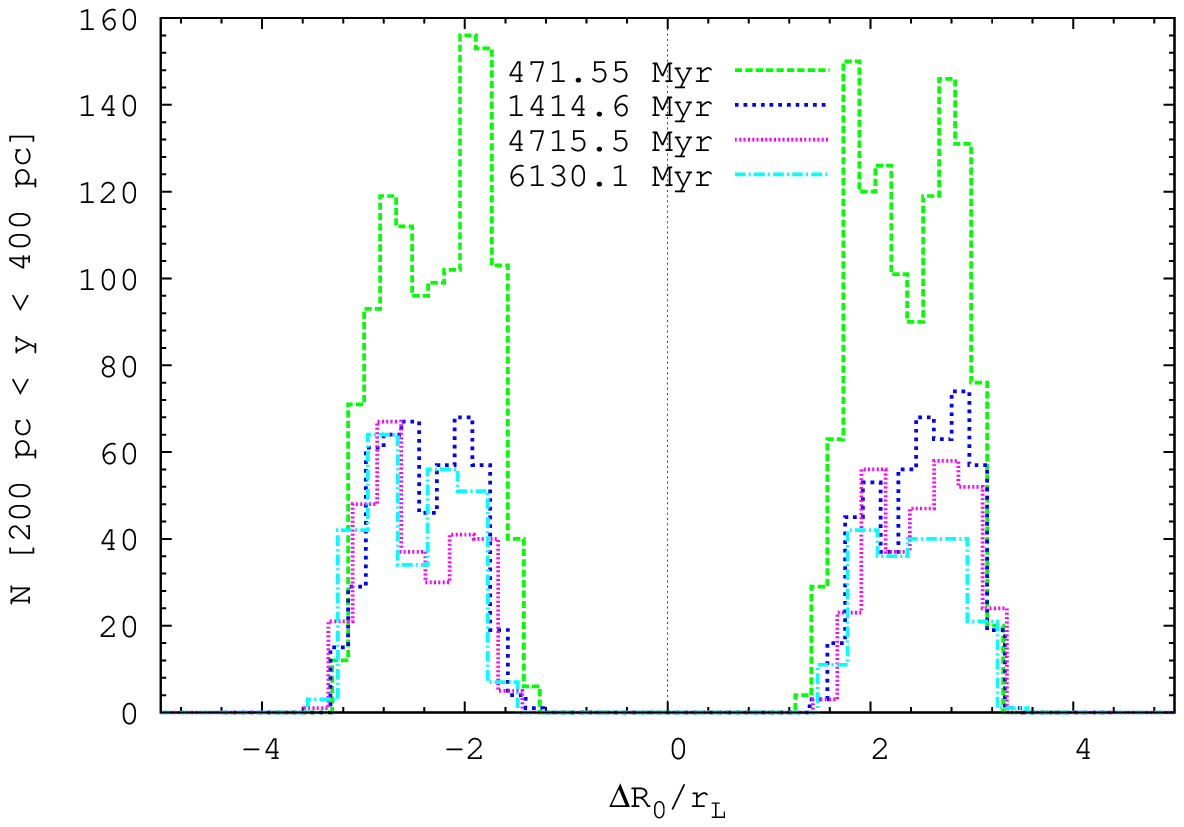}
  \includegraphics[width=0.45\textwidth]{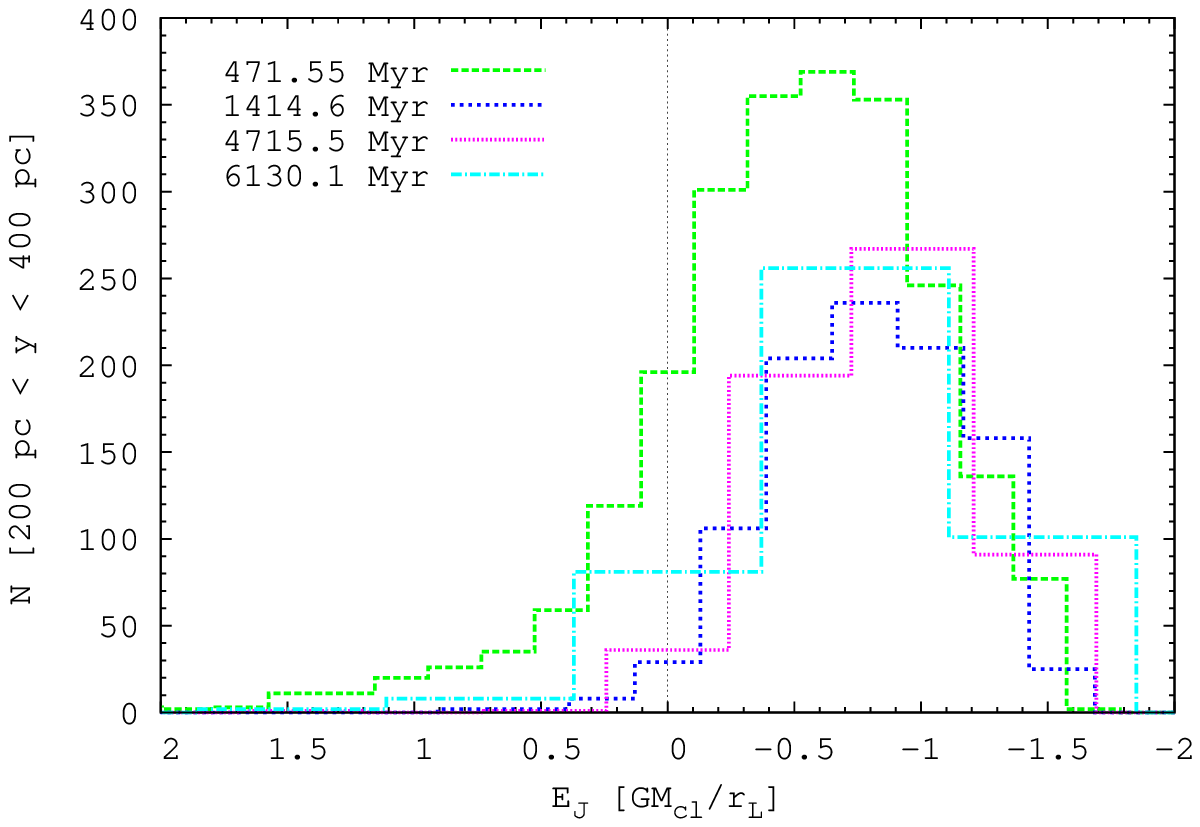}
  \end{center}
  \caption{Top: The histograms show the distribution of the radial offsets
  $N(\Delta R_0)$ of model 10 at different evolution times normalized to the
  corresponding tidal radii $r_\mathrm{L}(t)$.
  Stars at distances between 200\,pc$<|y|<$400\,pc along the tidal arms were
  selected. 
  Bottom: The Jacobi energy distribution $N(E_\mathrm{J})$ of the same stars 
  normalized to the actual $GM_\mathrm{cl}(t)/r_\mathrm{L}(t)$.
  Note that $N$ is the number of particles per bin and the bin width varies due
  to the different scaling.
  }
  \label{R0-evol}
\end{figure}

The radial offset $\Delta R_0$ of the epicentres of the tidal tail stars is
predicted by equation~\ref{eq-dR0} from the measured $\Delta L$.
In Figure~\ref{R0-evol} the histograms of calculated $\Delta R_0$ 
scaled to the corresponding
tidal radii $r_\mathrm{L}$ are shown for different times. For the time resolution we have
selected stars at tangential distances 200\,pc$<|y|<$400\,pc. We find that 
$\Delta R_0$ is proportional to $r_\mathrm{L}$ which scales with $M_\mathrm{cl}^{1/3}$. In the
next section we quantify the scaling. The corresponding distributions of Jacobi
energies $E_\mathrm{J}$ are shown in the lower panel of Figure~\ref{R0-evol}. Here
$E_\mathrm{J}$ is normalized to the actual $GM_\mathrm{cl}(t)/r_\mathrm{L}(t)$.

%----------------------------------------------------------------------------%
%----------------------------------------------------------------------------%

\begin{figure}
  \begin{center}
% \vspace{-0.2\textwidth}
  \includegraphics[width=0.45\textwidth]{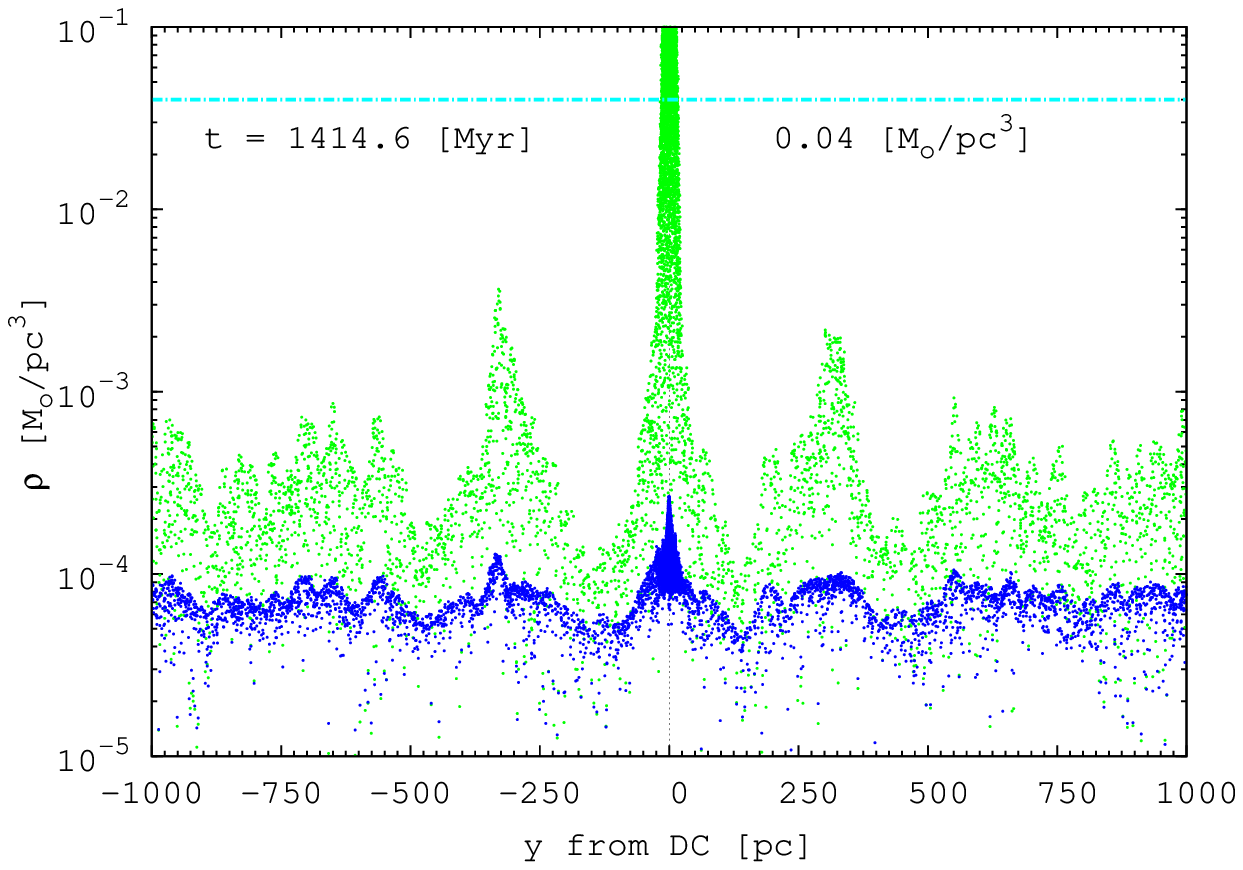}
  \includegraphics[width=0.45\textwidth]{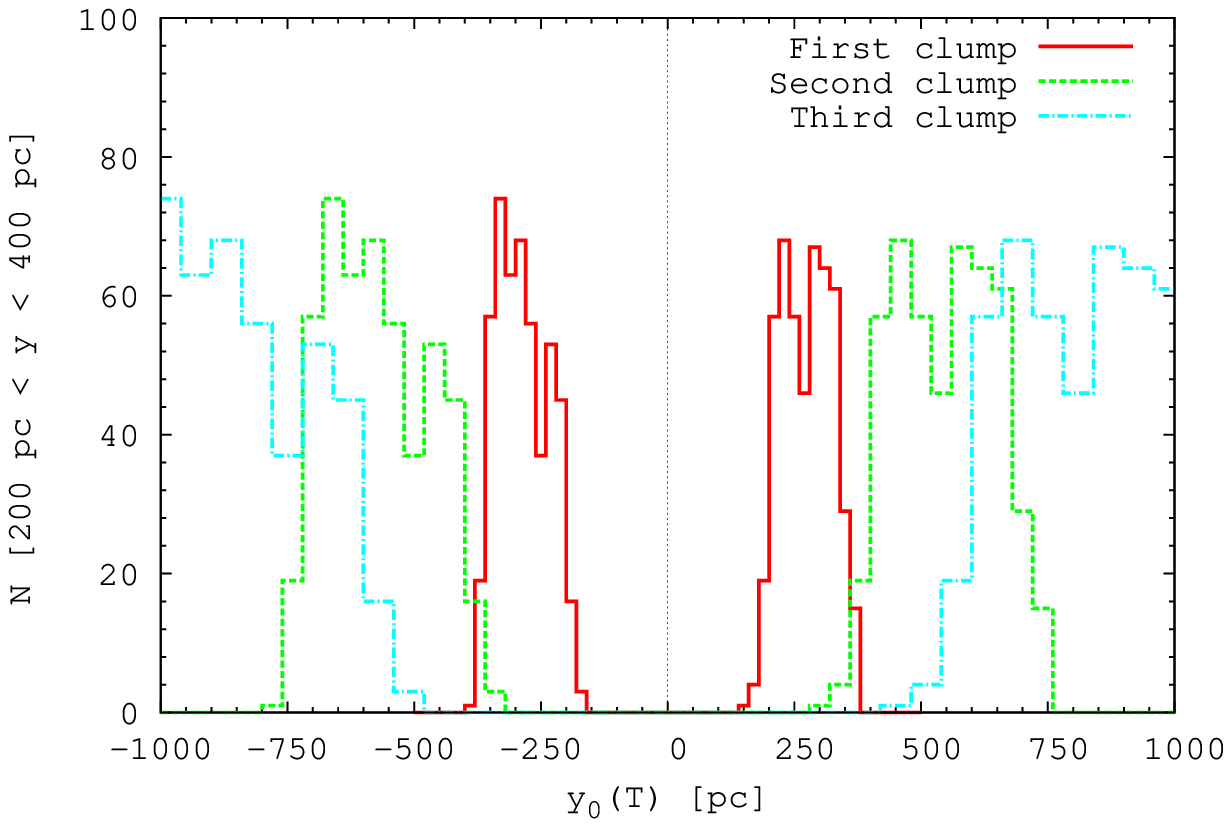}
  \end{center}
  \caption{Density and $y_0(T)$ plot along the tidal tails at t=1.4\,Gyr of 
  model 10. Grey (green) dots are
  the local densities and black (blue) dots show the mean density averaged 
  over a 30\,pc sphere. 
  The horizontal line corresponds to the stellar density in the solar
  neighbourhood.}
  \label{den-0300}
\end{figure}

\begin{figure}
  \begin{center}
% \vspace{-0.2\textwidth}
  \includegraphics[width=0.45\textwidth]{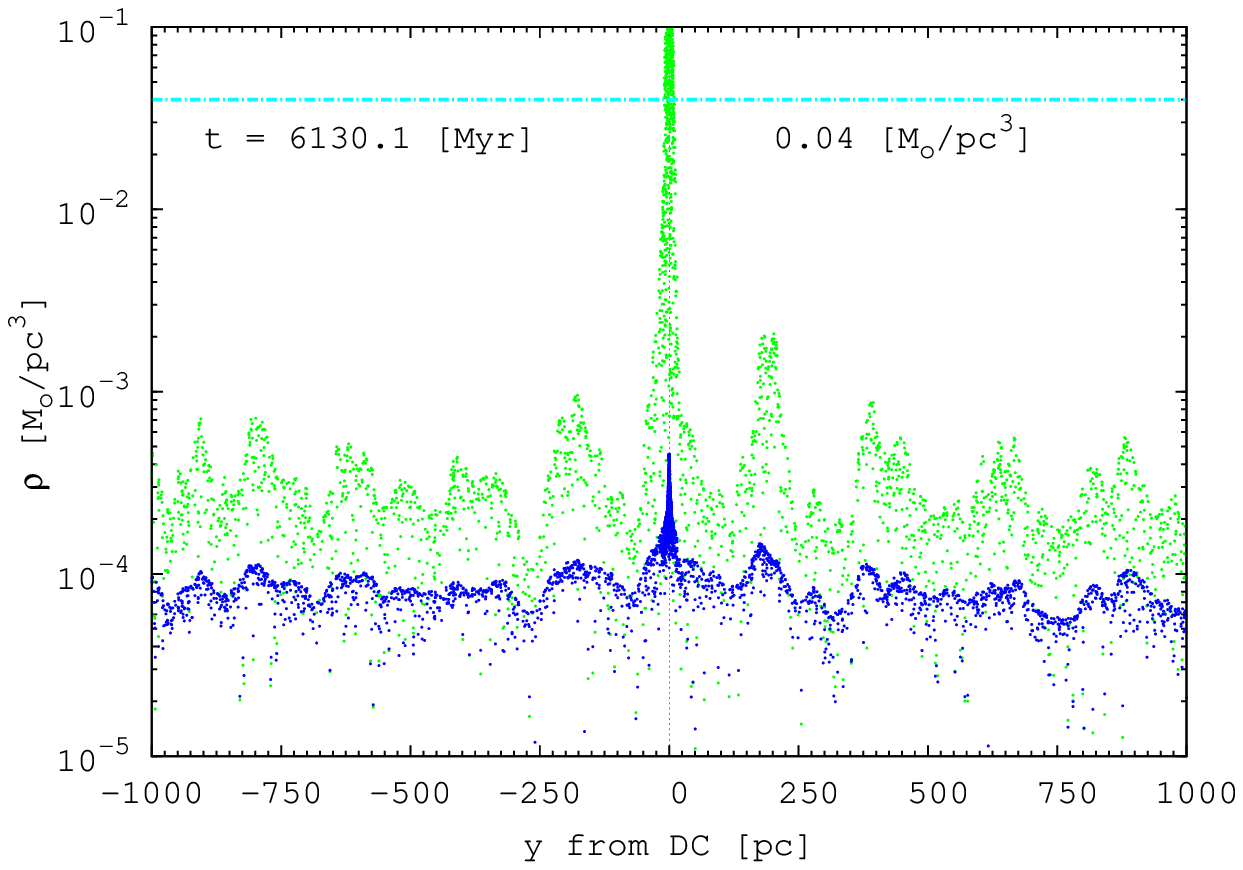}
  \includegraphics[width=0.45\textwidth]{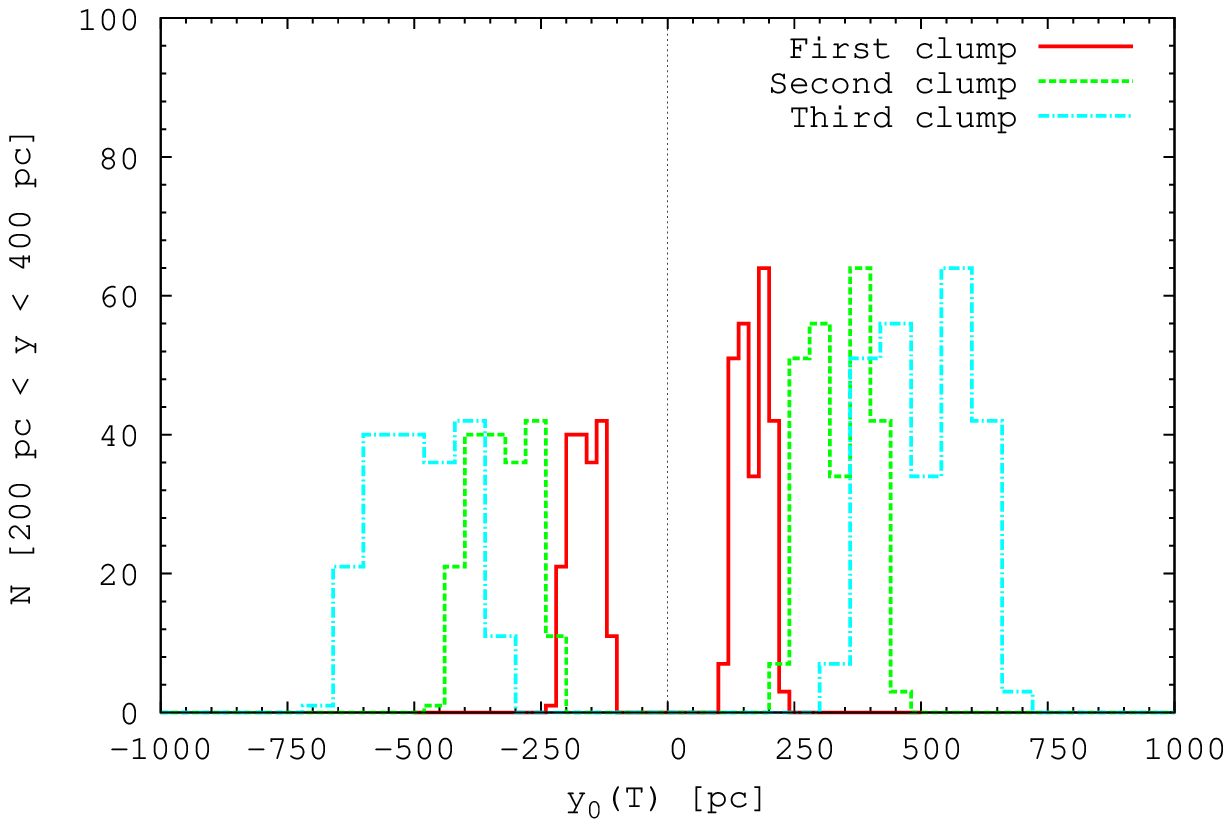}
  \end{center}
  \caption{Same as in Figure~\ref{den-0300} at time t=6.1\,Gyr.}
  \label{den-1300}
\end{figure}

The tangential positions of the clumps are multiples of $y_0(T)$, which are
connected to $\Delta L$ (equation~\ref{eq-y0-dl}).
Figures~\ref{den-0300} and \ref{den-1300} show a comparison of the 
density distribution along the tidal tails and predicted histograms of 
$1,2,3\times y_0(T)$ for the early and late time. The position and width of 
the clumps agree well in both plots and the second and third clump show
some overlap as predicted. 

%----------------------------------------------------------------------------%
%----------------------------------------------------------------------------%

%----------------------------------------------------------------------------%
%----------------------------------------------------------------------------%

A similar comparison of the apo- and pericentre distribution in 
the radial coordinate of the numerical simulation also matches the 
spread of the analytic prediction.

\begin{figure}
  \begin{center}
% \vspace{-0.2\textwidth}
  \includegraphics[width=0.45\textwidth]{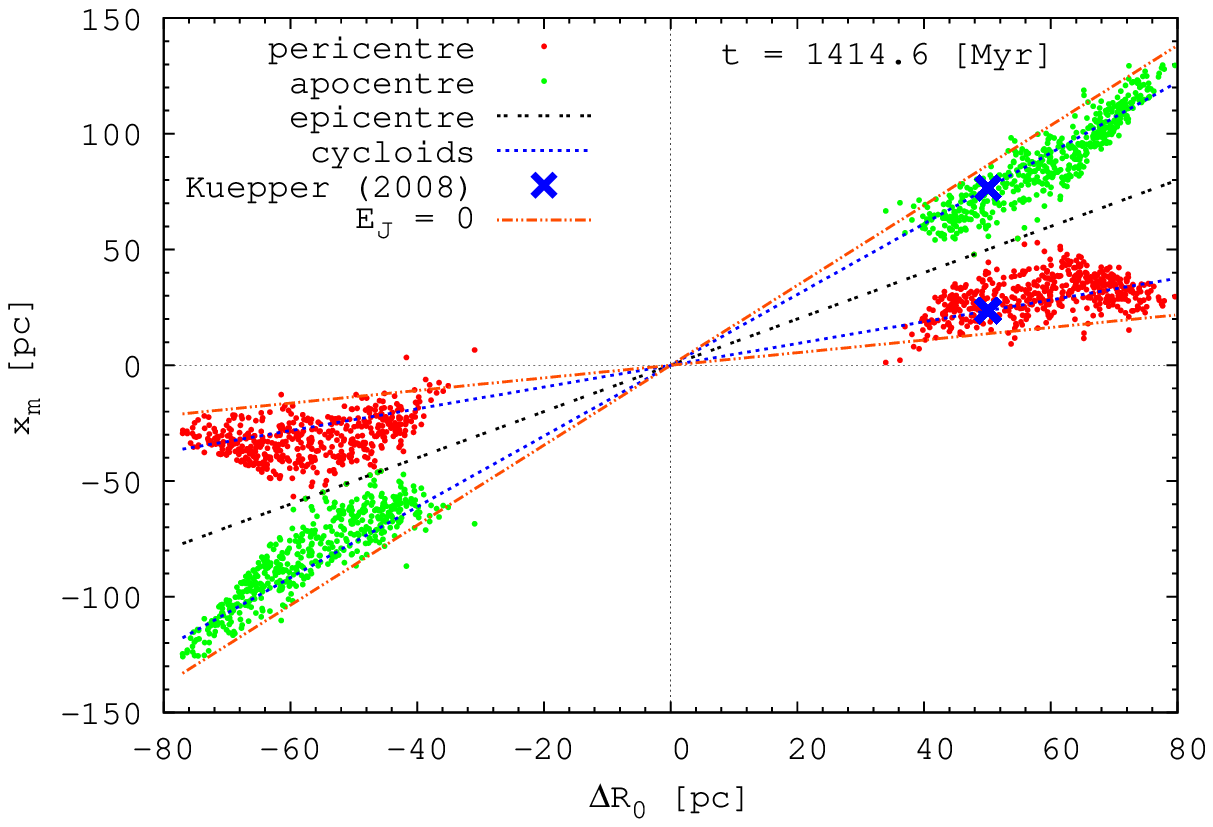}
  \end{center}
  \caption{Apo- and pericentre positions of tidal tail stars of model 10 as 
  function of $\Delta R_0$. 
  The double-dotted (black) line shows the epicentre position, 
  the dotted (blue) lines
  are the peri- and apocentre positions with zero velocity at pericentre
   in the corotating frame, and the 
   dot-dashed (orange) lines are peri- and apocentre positions 
   with $E_\mathrm{J}=0$. The cross marks the orbital parameters adopted in 
   \citet{Ku08}.}
  \label{rapo-300}
\end{figure}

In Figure~\ref{rapo-300} the peri- and apocentre positions of the selected
stars are shown as function of $\Delta R_0$. There is a strong correlation
between epicentre offset $\Delta R_0$ and amplitude, that is of 
$\Delta L$ and $\Delta E$. The double dotted (black) line shows
the epicentre position and the dotted (blue) lines show the pericentre position
 with
$v_\mathrm{t}=0$ and the corresponding apocentre (cycloids in the corotating 
reference
frame). The dot-dashed (orange) lines are for $E_\mathrm{J}=0$ which determines
 the maximum amplitude for most stars. Only very few stars fall outside this
limit.
The orbit adopted by \citet{Ku08} with ($x,v$)=($r_\mathrm{L},0$) is marked by 
the (blue)
crosses. It is a typical orbit with a slightly smaller epicentre offset 
$\Delta R_0$ compared to the mean value. Therefore they underestimated 
the distance of the clumps slightly as was already obvious from their simple 
$N$-body simulation.

%----------------------------------------------------------------------------%
%----------------------------------------------------------------------------%

%%%%%%%%%%%%%%%%%%%%%%%%%%%%%%%%%%%%%%%%%%%%%%%%%%%%%%%%%%%%%%%%%%%%%%%%%%%%%

\subsection{Parameter variation}
\label{sec-para}

In models 01-09 of Table~\ref{model-list} we vary the cluster mass 
and the galactocentric distance. Model 10 at an exact circular orbit differs
in the evolution of model 08 only in a small modulation of the mass loss.
Therefore we discuss here models 01-09 only. In the Section~\ref{sec-gc} a 
model near the galactic centre is discussed to show that the theory holds also
 in this extreme case.

\begin{figure}
  \begin{center}
% \vspace{-0.2\textwidth}
  \includegraphics[width=0.45\textwidth]{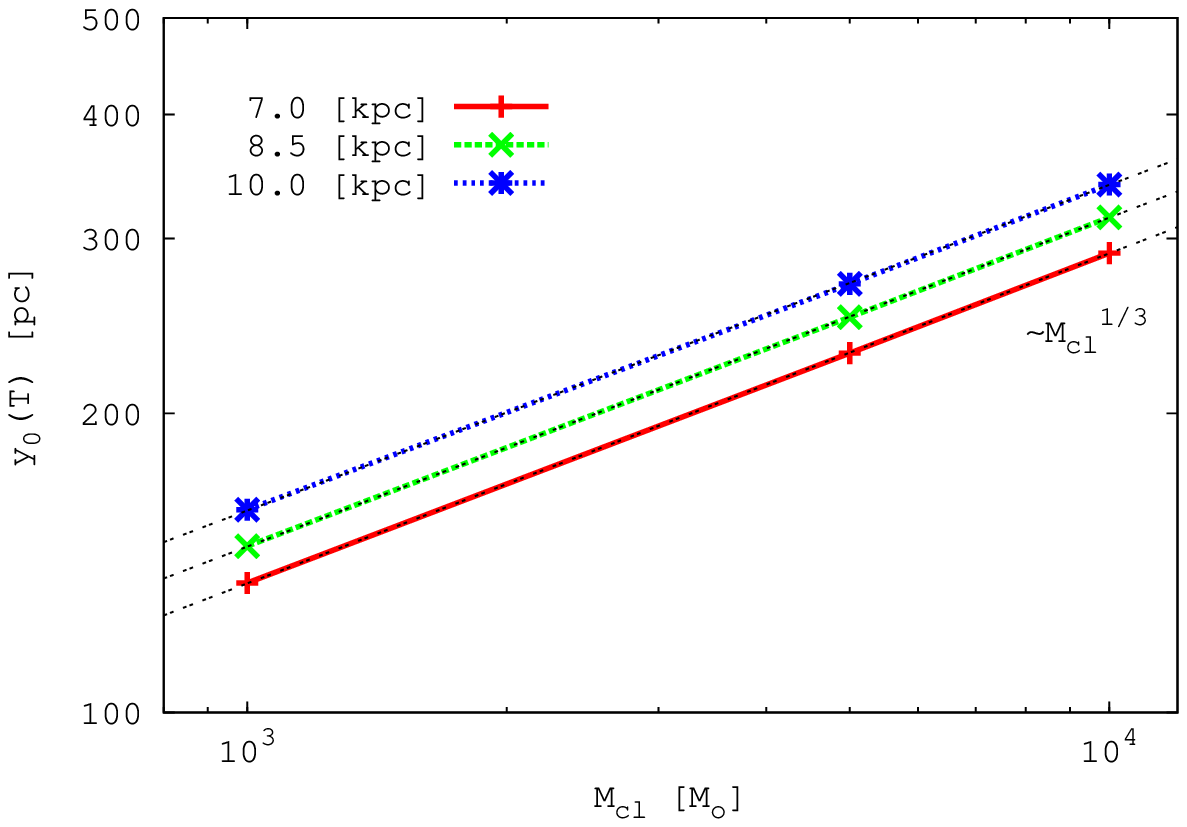}
  \includegraphics[width=0.45\textwidth]{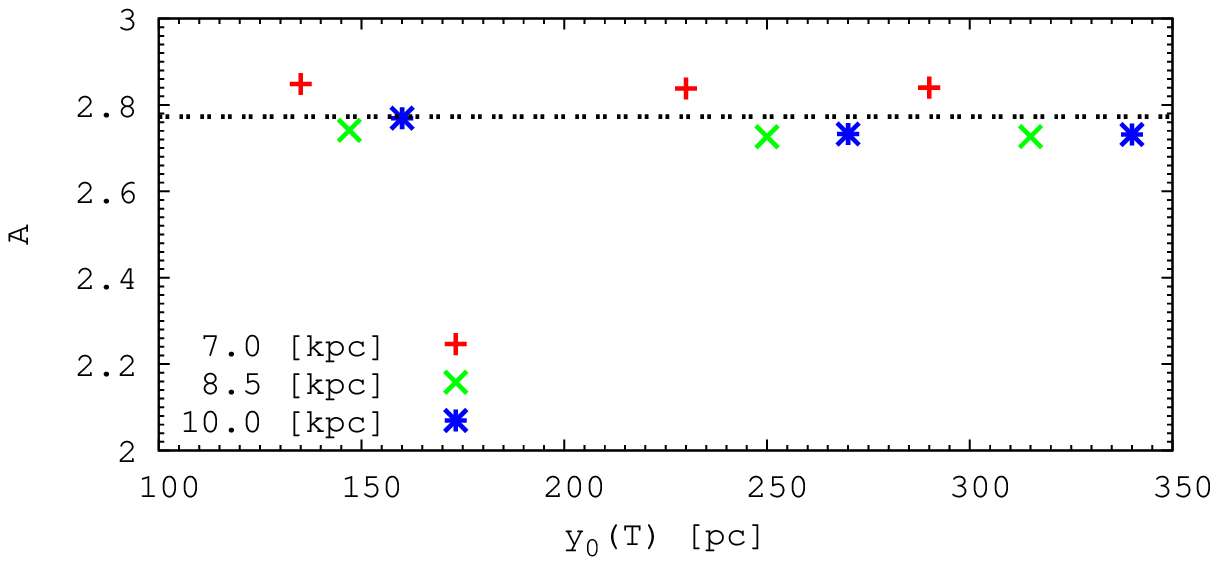}
  \end{center}
  \caption{Top: Mass dependence of $y_0(T)$ for the nine different models.
  The lines show power law best fits of $y_0$. The
  points and lines are sampled by galactocentric distance.
  Bottom: Scaling factor $A=\Delta R_0/r_\mathrm{L}$ for models 01-09 with the
  same coding as above and the best fit value for $A$.
}
  \label{res-tot}
\end{figure}

Already in the temporal evolution of model 10 we found that the tidal tail
structure scales with the tidal radius $r_\mathrm{L}$.  
For testing the dependence of $\Delta R_0$ and $y_0(T)$ on the cluster mass
$M_\mathrm{cl}$ and on the distance $R_\mathrm{C}$ to the Galactic centre we 
measure for all nine models
 the value of $y_0(T)$
numerically. We add the trailing and leading arm density distributions and
determine $y_0(T)$ by fitting the positions of clumps 1 and 2 to  
the value of 1$\times y_0(T)$ and 2$\times y_0(T)$. The result is plotted in
Figure~\ref{res-tot}. The upper panel shows the mass dependence. Best power law
fits for each $R_\mathrm{C}$ give a power law index of 1/3 to better than 1\%. 
We test the scaling of $\Delta R_0$ with $r_\mathrm{L}$ by defining
a scaling factor $A$
\bq
\Delta R_0 = A\, r_\mathrm{L} \label{Am}
\eq
and calculating $A$ from $y_0(T)$ using equation~\ref{eq-y0-dR0} 
\bq
A=\frac{1}{\pi}\frac{\beta}{4-\beta^2}\frac{y_0(T)}{r_\mathrm{L}} \label{At}
\eq
The result for the nine models is shown in the lower panel of Figure~\ref{res-tot}. The values of
$A$ are independent of $y_0(T)$, that is of $M_\mathrm{cl}$ and $R_\mathrm{C}$.
 The best fit
value is $A=2.77 \pm 0.02$. A possible dependence on $\beta$ cannot be tested
here, because the $\beta$ values are very similar. A further discussion is given
in Section~\ref{sec-gc}.

%----------------------------------------------------------------------------%
%----------------------------------------------------------------------------%

%%%%%%%%%%%%%%%%%%%%%%%%%%%%%%%%%%%%%%%%%%%%%%%%%%%%%%%%%%%%%%%%%%%%%%%%%%%%%

\subsection{The Galactic centre case}
\label{sec-gc}

%----------------------------------------------------------------------------%
\begin{table*}
%\begin{table*}[htbp]
\caption{The list of tidal clump parameters for the Galactic centre case as described in the text.}
\label{gc-result}
%\begin{flushleft}
\begin{center}
\begin{tabular}{llrrrrrrrrrr}
\hline\noalign{\smallskip}
 Arm & Clump & $\Delta L/L_\mathrm{C}$ & $\varphi_0$ [deg.] & $\varphi_\mathrm{L1} [deg.]$ &
 $\varphi_\mathrm{L2}$ [deg.] & $\Delta\varphi/\varphi_0$ [\%] & $r_\mathrm{L}$ [pc] & $y_0$ [pc] & $A_\mathrm{\varphi}$ & $A_\mathrm{L2}$ & $\Delta A/A_\mathrm{L2}$ [\%]\\
\noalign{\smallskip}
\hline
\noalign{\smallskip}
Leading & 1& -0.2179 & 51.6 & 43.3 & 44.1 & 14.5 & 2.45 & 16.9 & 1.81 & 1.54 & 14.9 \\
        & 2 & -0.2242 & 55.3 & 44.5 & 45.4 & 17.9 & 2.76 & 18.1 & 1.72 & 1.40 &	18.6 \\
Trailing & 1 & 0.2835 & 65.8 & 56.3 & 54.8 & 16.7 & 2.45 & 21.6 & 2.31 & 1.95 & 15.6 \\
%        & 2 & 0.3965 & 83.2 & 78.7 & 75.9 & 8.8 & 2.76 & 27.9 & 2.65 & 2.41 & 9.0 \\ 
\hline
\end{tabular}
%\end{flushleft}
\end{center}
\end{table*}
%----------------------------------------------------------------------------%

Near the Galactic centre the tidal forces are much stronger and the size 
of star clusters relative to the distance to the Galactic centre is 
much larger. Therefore it is an interesting case to test the epicyclic 
approximation in this extreme regime. We use the code {\sc nbody6gc}
to simulate the evolution of a star cluster in the tidal field of the
Galactic centre. This code is based on the parallel $N$-body code 
{\sc nbody6++} \citep{Aar1999,Aar2003,Spu1999} and in detail described 
in \citet{E08}. The orbits of the stars in the star
cluster are followed with a $4$th-order Hermite scheme \citep{Mak1992} 
including Kustaanheimo-Stiefel regularization of close
encounters \citep{Kus1965} and Chain regularization
\citep{Mik1998}. In addition, 
the orbit of the star cluster in the analytic background potential of 
the galactic centre is followed using an 
$8$th-order composition scheme \citep{Yos1990, McLa1995} including the
Chandrasekhar dynamical friction force with a variable Coulomb logarithm 
\citep{Ju05}. The dissipative force is numerically implemented
with an implicit midpoint method \citep{Mik2002}.

%----------------------------------------------------------------------------%
\begin{figure}
  \begin{center}
\hspace{-0.3cm}
  \includegraphics[width=0.55\textwidth]{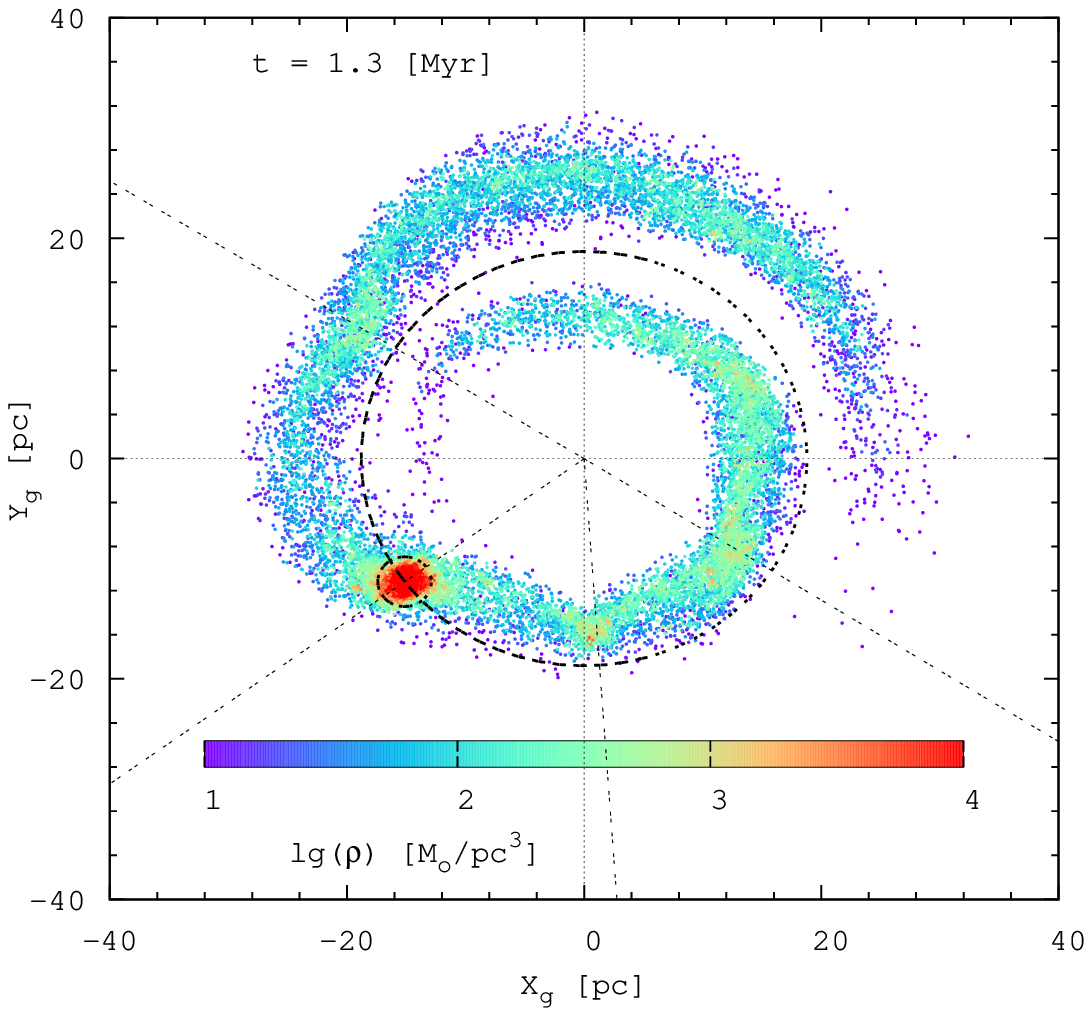}
 \end{center}
  \caption{Clumps in the tidal tails of a star cluster with $N=10^5$ 
  particles near the Galactic centre at $t = 1.3$ Myr.
  The color coding shows the local stellar density in logarithmic scale.
  The solid lines show the tidal radius and assigns the galactocentric distance
  of the cluster. The radial lines mark the angles of the first and second 
  clump of the leading arm and the first clump of the trailing arm with respect 
  to the cluster centre.}
  \label{knots130}
\end{figure}
%----------------------------------------------------------------------------%

%----------------------------------------------------------------------------%
\begin{figure}
  \begin{center}
% \vspace{-0.2\textwidth}
  \includegraphics[width=0.45\textwidth]{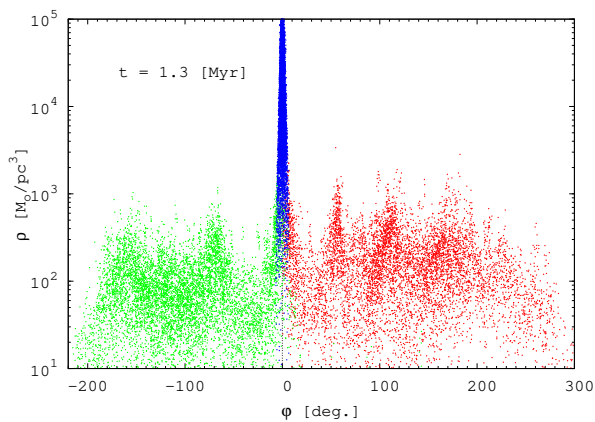}
 \end{center}
  \caption{Density along the tidal arms as a function of
  the angle with respect to the cluster centre. The density maxima
  can be clearly identified. Stars inside the tidal radius corresponding
  to bound stars are in black (blue), the leading
  arm in dark grey at positive angles (red), and the trailing arm in 
  light grey at negative angles (green). The wrap
  of the tidal tails can be seen from the $\varphi$-range exceeding 360$^0$.}
  \label{den-andreas}
\end{figure}
%----------------------------------------------------------------------------%

%----------------------------------------------------------------------------%
\begin{figure}
  \begin{center}
% \vspace{-0.2\textwidth}
  \includegraphics[width=0.45\textwidth]{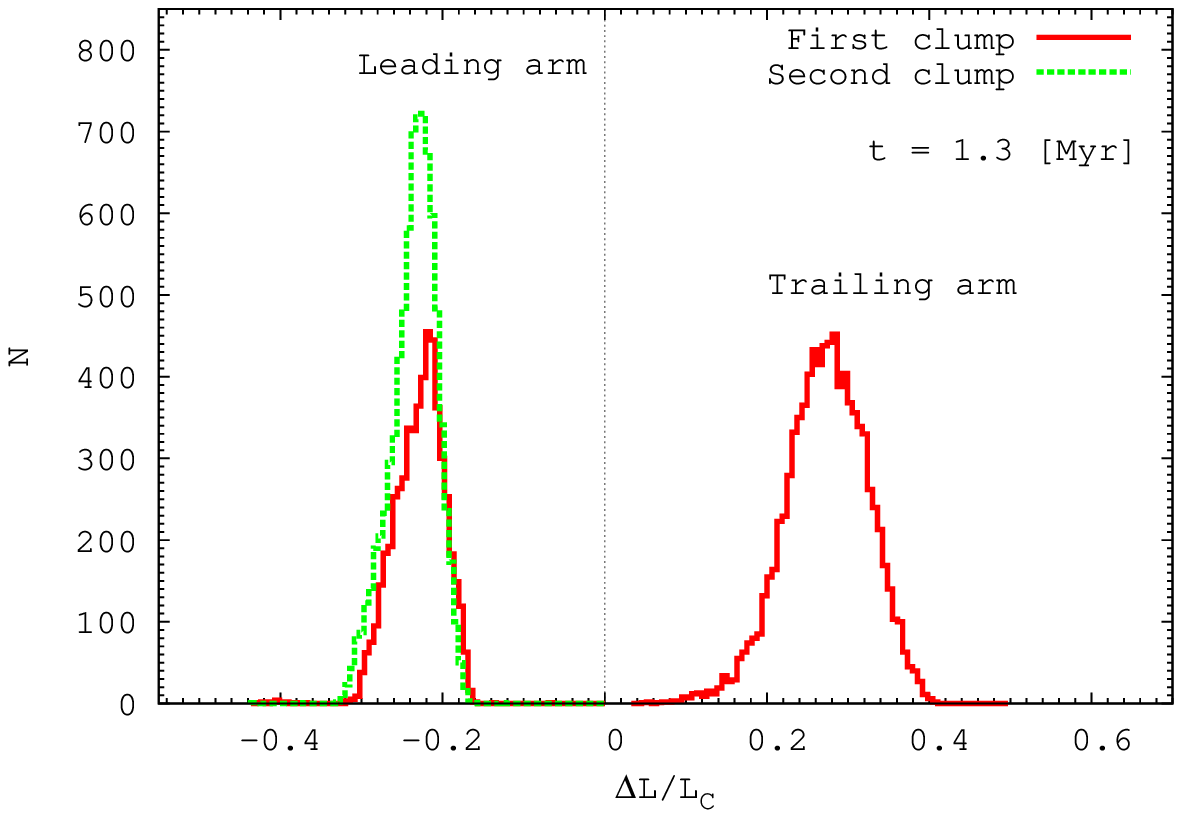}
 \end{center}
  \caption{Histogram of angular momentum differences 
  $\Delta L/L_\mathrm{C}$ scaled by the angular momentum of the circular orbit. 
  $N$ is the number
  of particles with a certain angular momentum difference. Only stars
  within 25 degrees around the density maxima in the clumps have been 
  included in the statistics.}
  \label{dll}
\end{figure}
%----------------------------------------------------------------------------%

For the Galactic centre, we used a scale free model (i.e., with a 
power law density profile, e.g. \citet{MDZ1996}) with a 
supermassive black hole \citep{Eis2005} added at the centre.
 The cumulative mass profile is given by 
 $M(R)= M_\mathrm{bh} + M_0 (R/R_0)^\alpha$.
The parameters were taken to roughly match those in the centre of
the Milky Way. We used $M_\mathrm{bh}=3.6\times 10^6 M_\odot$, 
$M_0=1.67\times 10^8 M_\odot$ at $R_0=20$\,pc
and $\alpha=1.2$.
At a distance of $R_\mathrm{C}$=20\,pc the influence of the central black
hole can be neglected and the model is scale free.
In the limit of a scale free model, the ratio
$\beta=\kappa/\Omega$ is independent of
galactocentric distance and is given by
$\beta_\mathrm{S}^2 = \alpha + 1 = 2.2$.
For the star cluster, we used a King model \citep{Ki1966}
with $W_0=6$, mass of $M_{\rm cl}=10^6 M_\odot$, and 
half-mass radius of $r_\mathrm{h}=1.64$\,pc starting
at a galactocentric radius of $R_\mathrm{C}=20$\,pc. 
The particle number is $N=10^5$.
The initial tidal radius is $r_\mathrm{L}=2.72$\,pc (mean of L1 and L2).

Figure \ref{knots130} shows the tidal arms for the simulation after an 
evolution time of $t$=1.3\,Myr. 
At $t$=1.3\,Myr the galactocentric distance has slowly decayed due to 
dynamical friction to $R_\mathrm{C}=18.8$\,pc and the tidal radius decayed 
mainly due to cluster mass loss to $r_\mathrm{L}=2.26$\,pc according to
equation \ref{eq-rL} (both marked by circles). 
The local density is colour coded showing clearly the density maxima in the
tidal tails. 
Three clumps can be identified in the leading arm and two clumps in the 
trailing arm. The radial lines from the Galactic centre mark the angles 
of three of these clumps with respect to the cluster centre.

Figure \ref{den-andreas} shows the density along the tidal arms
as a function of azimuth angle $\varphi$ with respect to the cluster centre.
The clumps can be clearly identified as peaks. The $\varphi$-range exceeds
360$^0$ showing the wrap of the tidal tails.

Figure \ref{dll} shows the histogram of the number of stars as function 
of angular momentum difference $\Delta L/L_\mathrm{C}$. We included only stars
within an angle of 25 degrees around the density maxima in the statistics. 
 The distribution is asymmetric with respect to the leading and trailing arms
which cannot be explained in the frame of the epicyclic theory.
 The maxima in the histogram correspond to the density maxima of the
clumps in the leading and trailing arms, respectively.

The measured angles $\varphi_0=y_0(T)/R_\mathrm{C}$ of the density maxima in the clumps can be 
compared to the theoretical 
estimates $\varphi_\mathrm{L}$ detrived from $\Delta L$. 
Note that these are the angles between the cluster centre 
and the first clump or the first and the second clump. For the theoretical estimate, 
we plugged the most frequent angular momentum differences of the leading and trailing 
arms from Figure \ref{dll} into equation

\begin{eqnarray}
\varphi &=& \frac{2\pi}{\beta_\mathrm{S}}\left[ 1 - \frac{\Omega_\mathrm{C}}{\Omega} \right] \\
&\simeq& \frac{2\pi}{\beta_\mathrm{S}}\left[ \frac{\beta_\mathrm{S}^2-4}{\beta_\mathrm{S}^2} \frac{\Delta L}{L_\mathrm{C}}
-\frac{\beta_\mathrm{S}^2-2}{\beta_\mathrm{S}^2}\frac{\beta_\mathrm{S}^2-4}{\beta_\mathrm{S}^2} \frac{\Delta L^2}{L_\mathrm{C}^2}\right].
\end{eqnarray}

\noindent
Compared to equation~\ref{eq-y0-dl} we added here the second order term to
test the
sensitivity of the results.
The results are shown in Table \ref{gc-result} for the first two clumps in
the
leading arm and the first clump in the trailing arm. The density maximum
of the second clump in the trailing arm is not well-defined.
We denoted the measured angle as $\varphi_0$. The theoretical estimates
from the measured $\Delta L/L_\mathrm{C}$ have been denoted as
$\varphi_\mathrm{L1}$ (first-order)
and $\varphi_\mathrm{L2}$ (second-order).
We defined $\Delta\varphi = \varphi_0-\varphi_\mathrm{L2}$
for the error. Note that $\varphi_\mathrm{L2}$ is systematically lower
than
$\varphi_0$,
but in principle we find a good agreement between measurement and theory.
We also calculate the A values $A_\mathrm{L2}$ and $A_\mathrm{\varphi}$
from equations (\ref{Am}) and (\ref{At}). $A_\mathrm{L2}$ is calculated
from the
Taylor expansion of $\Delta R_0$ in $L$ to second-order.
For $A_\mathrm{\varphi}$ we used the measured $y_0=R_\mathrm{C}\varphi_0$.
We used the tidal radius according to equation (\ref{eq-rL}) at the time
$t-cT$, where the stars in the clumps were released from the cluster.
Here $T$ is the epicyclic period at $R_\mathrm{C}$ and $c=1$ and $c=2$
correspond to the first and second clump, respectively.
We defined $\Delta A = A_\mathrm{\varphi}-A_\mathrm{L2}$ for the error.
Again, the
agreement is relatively good. However, the $A$ value is significantly
smaller than
that at large galactocentric distances.

There are some aspects of the tidal tail structure which
cannot be explained by the simple epicyclic theory.
The most prominent one is the asymmetry in the tidal arms
concerning the angular momentum and Jacobi energy distribution.
The errors $\Delta\varphi/\varphi_0$ and $\Delta A/A_{L2}$
stem from a slight non-conservation of angular momentum
in the tidal arms. The reason is the influence of the cluster
potential. This needs to be investigated further.
The asymmetry between the inner and outer Lagrange points with
respect to the central potential of the cluster is only a few percent.
However, due to the proximity to the Galactic centre, the phase space
for the particles which escape into the leading arm is considerably
smaller than for those which escape into the trailing arm.
The streaming velocitiy differs considerably between the leading
and trailing arms. Thus the redistribution of energy and angular
momentum for fixed $E_\mathrm{J}$ can be different. Since the radial
offset is not small compared to the distance to the Galactic
centre, an epicyclic theory for larger amplitudes would be
helpful. A Taylor expansion in $\eta=1/R$ holds up to eccentricities
of 0.5 as was shown by \citet{De76} (see also \citet{Ar06}
for a derivation in the solar neighbourhood).
For a further investigation we refer to \citet{E08}, where the effects
due to third- and higher-order terms in the Taylor expansion
of the effective potential are discussed.
%%%%%%%%%%%%%%%%%%%%%%%%%%%%%%%%%%%%%%%%%%%%%%%%%%%%%%%%%%%%%%%%%%%%%%%%%%%%%

\section{Summary}
\label{sec-sum}

We presented a quantitative derivation of the angular momentum 
and energy distribution of escaping stars from a star cluster in 
the tidal field of the Milky Way. Despite the motion on a circular 
orbit, the tidal tails are clumpy due to the epicyclic motion of 
the stars. We compared the derived distances and widths of the 
clumps with numerical simulations using star-by-star simulations. 
For star clusters at the solar circle we included an IMF and mass 
loss due to stellar evolution in the calculations. The same 
equations were applied to a star cluster very close to the 
Galactic centre, where the tidal forces are very strong.

We find a very good agreement of theory and models concerning the tidal tail
structure. The positions of the clumps are determined by the angular momentum
offset of the stars, which lead to a radial offset of the epicenters with
respect to the cluster orbit. 
The investigation of \citet{Ku08} is a special case of our 
investigations but for a Kepler potential. We find that the radial offset of 
the tidal arms is proportional to the tidal radius. However near the Galactic 
centre the factor of proportionality is considerably smaller. 

The tidal arm structure at large galactocentric radii is symmetric, 
whereas the asymmetry near the galactic centre is considerable. 
This can be reproduced only partly by taking into account the correction of the
epicyclic frequency at the epicentre radii.

We have also measured the Jacobi energy distribution of bound 
stars and showed that there are 35\% of stars above 
the critical Jacobi energy independent of the evolutionary state 
of the cluster. These stars can potentially leave the 
cluster. 
This is a hint, that mass loss is dominated by a self-regulating process of 
increasing Jacobi energy due to the diminishing gravitational 
potential of the star cluster induced by the mass loss itself.
 
Finally we consider the observability of the predicted clump properties 
in the tidal tails of star clusters. The identification of tidal tail stars of
open clusters on a circular orbit is strongly hampered by the large number of
nearby field stars with similar properties. But with differential methods using
high quality data for distances and velocities it may be possible to identify
the most prominent first clumps. 
We have shown that the maximum density in the first clump does
not decrease with time until dissolution of the cluster. The first clumps are
formed by escaping stars with a time delay determined by the epicyclic
period $T_\kappa\approx 150$\,Myr. The tidal tail structure will probably 
survive the gravitational
scattering process, which is also responsible for the galactic disc heating
\citep{Wie77}. On a timescale of one epicyclic period, 
we expect only small perturbations of the tidal clump position 
and velocity but no destruction.
The density maximum of the first clump is of the order of a few percent of the 
field density of the galactic disc. The velocity imprint by the epicyclic motion
is of the order of 2\,km/s. An overdensity with these properties
may be difficult to observe for clusters on exact circular orbits, if there is no
additional separating property. For young star clusters the high fraction of
early type stars can serve for such a discrimination.
On the other hand most star clusters are identified by the systematic peculiar
 motion with respect to the field stars. If the positions and velocities of the
 tidal clump stars are properly predicted, they may 
be observable as moving groups.

For an application of the tidal clump theory to the eccentric orbits of
globular clusters a perturbation theory with respect to the 'free falling'
comoving coordinate system would be necessary. In this case the Jacobi energy is
no longer a constant of motion. This will be a matter of future investigations.
Some numerical test runs have shown that tidal tail clumps are formed also on
highly eccentric orbits without an additional external perturbation, but the
geometry and density is modulated along the orbital position.

%%%%%%%%%%%%%%%%%%%%%%%%%%%%%%%%%%%%%%%%%%%%%%%%%%%%%%%%%%%%%%%%%%%%%%%%%%%%%

\section{ACKNOWLEDGEMENTS}

\textit{P. B. \& M. P.} thanks for the special support 
of his work by the Ukrainian National Academy of Sciences under 
the Main Astronomical Observatory GRAPE/GRID computing 
cluster project.

\textit{P. B.} acknowledges his support from the German 
Science Foundation (DGF) under SFB 439 (sub-project B11) 
at the University of Heidelberg. His work was also supported 
by the Volkswagen Foundation GRACE Project No. I80 041-043.

\textit{M. P.} acknowledges support by  the University of 
Vienna through the frame of the Initiative Kolleg (IK) 
''The Cosmic Matter Circuit'' I033-N and computing time 
on the Grape Cluster of the University of Vienna.

\textit{A. E.} would like to thank Rainer Spurzem, Ortwin Gerhard
and Kap-Soo Oh for the provision of an earlier version of
{\sc nbody6gc} and gratefully acknowledges support by the 
International Max Planck Research School (IMPRS) for Astronomy 
and Cosmic Physics at the University of Heidelberg.

We thank the DEISA Consortium ({\tt www.deisa.eu}), co-funded 
through EU FP6 projects RI-508830 and RI-031513, for support 
within the DEISA Extreme Computing Initiative and the 
Astrogrid-D, which links together the two GRAPE clusters
in Kiev and Heidelberg.

%%%%%%%%%%%%%%%%%%%%%%%%%%%%%%%%%%%%%%%%%%%%%%%%%%%%%%%%%%%%%%%%%%%%%%%%%%%%%

%%%%%%%%%%%%%%%%%%%%%%%%%%%%%%%%%%%%%%%%%%%%%%%%%%%%%%%%%%%%%%%%%%%%%%%%%%%%%

\appendix

\section{Taylor expansions}
\label{app-epi}

We use Taylor expansions of the radial variation of different quantities in the
gravitational field of the Galaxy in $R$ and $L$ with respect to some
circular orbit with $R_0$ and $L_0$. The potential $\Phi_\mathrm{cl}$ of the star
cluster is not expanded. The Taylor expansions are applied to: 
1) the energy of stars on eccentric orbits with fixed $L$ in the galactic
field;
2) the effective potential in the cluster frame;
3) the epicentre position and energy excess of tidal tail stars in the cluster
frame.
For the Taylor expansions we use
\bqn
\kappa^2&=&2\Omega^2\left(2+\frac{\dd\ln \Omega}{\dd\ln R}\right)
	\label{eq-kappa}\\
\beta=\frac{\kappa}{\Omega} &\mathrm{and}& 
\beta'=\frac{\dd \beta}{\dd \ln R}\\ 
\frac{\dd \Phi_\mathrm{g}}{\dd R}=\Omega^2 R &\mathrm{and}& 
	\frac{\dd\ln \Omega}{\dd\ln R}=\frac{\beta^2-4}{2}
\eqn
with angular speed $\Omega$ and epicyclic frequency $\kappa$.

For the epicyclic motion in the tidal tails (Sec. \ref{sec-epi} and \ref{sec-gc}) 
the gravitational potential of the galaxy is needed in terms of $r=R-R_0$. 
We find to third order
\bqn
\Phi_\mathrm{g}(R)&=&\Phi_\mathrm{g}(R_0)+\frac{L_0^2}{R_0^3} r
	+\left(\beta_0^2-3\right)\frac{L_0^2}{R_0^4} \frac{r^2}{2}\\&&
+\left[(\beta_0^2-3)(\beta_0^2-4)+2\beta_0\beta'_0\right] 
	\frac{L_0^2}{R_0^5} \frac{r^3}{6}
\eqn

For the apo- and pericentre $R_\mathrm{m}=R_0+r_\mathrm{m}$ we find
for the kinetic energy
\bqn
\frac{L_0^2}{2R_\mathrm{m}^2}&=&\frac{L_0^2}{2R_0^2}
	\left(1-2\frac{r_\mathrm{m}}{R_0}+3\frac{r_\mathrm{m}^2}{R_0^2}-4\frac{r_\mathrm{m}^3}{R_0^3}\right)
\eqn
leading to the energy excess $\Delta E=E-E_0$ (relative to the circular orbit
with the same angular momentum $L_0$)
\bqn
\Delta E&=&\frac{ \beta_0^2}{2} \frac{L_0^2}{R_0^4}r_\mathrm{m}^2
+\left[\beta_0^2(\beta_0^2-7)+2\beta_0\beta'_0\right] 
	\frac{L_0^2}{R_0^5} \frac{r_\mathrm{m}^3}{6}
\eqn

For the derivation in the rest frame of the star cluster we need the Taylor
expansions of $R(L)$, $\Omega(L)$ and
$\Phi_\mathrm{g}(L)$ with respect to $\Delta L=L-L_\mathrm{C}$ ($L$ is the angular momentum 
of the circular orbit at radius $R(L)$).
In terms of position $x$ and velocity $v_\mathrm{r},v_\mathrm{t}$ in the corotating frame we have
\bq
\Delta L=(v_\mathrm{t}+\Omega_\mathrm{C} R)R-\Omega_\mathrm{C} R_\mathrm{C}^2
	=(2\Omega_\mathrm{C} x+ v_\mathrm{t}) R_\mathrm{C} +\Omega_\mathrm{C} x^2 + v_\mathrm{t} x
\eq
The Taylor expansions in $L$ are
\bqn
R(L)&=&R_\mathrm{C}+\frac{2}{\beta_\mathrm{C}^2}\frac{R_\mathrm{C}}{L_\mathrm{C}} \Delta L \label{eq-RofL}\\&&
	+\frac{1}{\beta_\mathrm{C}^4}\left[(2-\beta_\mathrm{C}^2)-\frac{4\beta'}{\beta}\right]
	\frac{R_\mathrm{C}}{L_\mathrm{C}^2} \Delta L^2\nonumber\\
\Phi_\mathrm{g}(L)&=&\Phi_{g,C}+\frac{2}{\beta_\mathrm{C}^2}\Omega_\mathrm{C} \Delta L \label{eq-phiofL}\\&&
+\frac{1}{\beta_\mathrm{C}^4}\left(\beta_\mathrm{C}^2-4-2\frac{\beta'_\mathrm{C}}{\beta_\mathrm{C}}\right)\Omega_\mathrm{C} \frac{\Delta L^2}{L_\mathrm{C}}
\nonumber\\
\frac{\Omega(L) L}{2}&=&\frac{\Omega_\mathrm{C} L_\mathrm{C}}{2}
	+\frac{\beta_\mathrm{C}^2-2}{\beta_\mathrm{C}^2}\Omega_\mathrm{C} \Delta L \label{eq-kinofL}\\&&
+\frac{1}{2\beta_\mathrm{C}^4}\left((\beta_\mathrm{C}^2-2)(\beta_\mathrm{C}^2-4)+4\frac{\beta'_\mathrm{C}}{\beta_\mathrm{C}}\right)\Omega_\mathrm{C} \frac{\Delta L^2}{L_\mathrm{C}}
\nonumber\\
\frac{1}{\Omega(L)} &=& \frac{1}{\Omega_\mathrm{C}} - \frac{\beta_\mathrm{C}^2-4}{\beta_\mathrm{C}^2}\frac{1}{\Omega_\mathrm{C}}\frac{\Delta L}{L_\mathrm{C}} \\
&&+ \frac{1}{\beta_\mathrm{C}^4} \left( (\beta_\mathrm{C}^2-2)(\beta_\mathrm{C}^2-4) - 8\frac{\beta_\mathrm{C}'}{\beta_\mathrm{C}}\right) \frac{1}{\Omega_\mathrm{C}}\frac{\Delta L^2}{L_\mathrm{C}^2}
\nonumber
\eqn
Here already second order terms contain $\beta'$.
Since $\beta\beta'=\dd^2 \ln\Omega/\dd (\ln R)^2$, 
the logarithmic derivative $\beta'$ vanishes only, if 
$\Omega(R)\propto R^{\alpha}$ is 
exactly a power law with constant $\alpha$. For realistic rotation
curves $\beta$ varies considerably (see Fig.~\ref{fig-rot}).
We find the energy excess $\Delta E_0=E_0-E_\mathrm{C}$ of the epicentre motion
for equation~\ref{eq-de0} by adding equations~\ref{eq-phiofL} and \ref{eq-kinofL}
\bqn
\Delta E_0&=&\Omega_\mathrm{C} \Delta L
	+\frac{1}{2\beta_\mathrm{C}^2}\left(\beta_\mathrm{C}^2-4\right)
	\Omega_\mathrm{C} \frac{\Delta L^2}{L_\mathrm{C}} \label{eq-dE0ofL}
\eqn
Here the $\beta'$ term vanishes.

For the derivation in the rest frame of the star cluster we need $\Phi_\mathrm{eff}$ only in terms of $r=R-R_\mathrm{C}$. The cluster potential is not expanded in a Taylor series. We find
\bqn
\Phi_\mathrm{eff}(R)&=&\Phi_\mathrm{cl}+\Phi_\mathrm{g,eff}(R_\mathrm{C})+\frac{\beta_\mathrm{C}^2-4}{2}\Omega_\mathrm{C}^2 r^2+ \nonumber \\&&
\left[(\beta_\mathrm{C}^2-3)(\beta_\mathrm{C}^2-4)+2\beta_\mathrm{C}\beta'_\mathrm{C}\right] \Omega_\mathrm{C}^2 \frac{r^3}{6R_\mathrm{C}}
\eqn
%%%%%%%%%%%%%%%%%%%%%%%%%%%%%%%%%%%%%%%%%%%%%%%%%%%%%%%%%%%%%%%%%%%%%%%%%%%%%

%%%%%%%%%%%%%%%%%%%%%%%%%%%%%%%%%%%%%%%%%%%%%%%%%%%%%%%%%%%%%%%%%%%%%%%%%%%%%
\label{lastpage}

\end{document}